\documentclass[a4paper,12pt]{article}
\usepackage[cmex10]{amsmath}
\usepackage{latexsym,amssymb}
\usepackage[mathscr]{eucal}
\usepackage{amsxtra,amscd,amsthm,upref,layout,color}
\usepackage{calc}
\usepackage[final]{graphicx}
\def\refup#1{{$^{#1}$}}

\def\mn{{\langle n\rangle}}\def\meta{{\langle \eta \rangle}}

\def\br{{\bf r}}\def\bq{{\bf q}}\def\ba{{\bf a}}\def\bb{{\bf b}}

\def\oo{{\leavevmode\setbox0=\hbox{h}\dimen0=\ht0 \advance\dimen0
by-1ex\rlap{\raise0.47\dimen0\hbox{\char'27}}o}}
\def\begeq{\begin{equation}}
\def\endeq{\end{equation}}
\def\begdis{\begin{displaymath}}
\def\enddis{\end{displaymath}}

\DeclareMathOperator*{\limM}{\rm{l.i.m.}}
\def\bA{{\bf A}}\def\bB{{\bf B}}\def\bC{{\bf C}}
\def\bSa{{\bf S$_1$}}\def\bSb{{\bf S$_2$}}
\def\cA{{\cal A}}\def\cD{{\cal D}}
  \def\cI{{\cal I}}
  
\def\cP{{\cal P}}  

 \def\cW{{\cal W}}\def\cWp{{{\cal W}_p}}\def\cWb{{{\cal W}_b}}

\def\S1{{S$_1$}}

\def\ie{{\em i.e.}}\def\eg{{\em e.g.}}
\def\etal{{\em et al.}}
\def\hw{{\hat \omega}}

\def\tn{{\tilde {n}}}

\def\teta{{\tilde {\eta}}}

\begin{document}
\title{On physical scattering density fluctuations of amorphous samples}
\author{ 
{{ 
 Salvino Ciccariello$^{a,b,*}$, Piero Riello$^{a}$ and Alvise Benedetti$^{a}$
}}\\
%
  \begin{minipage}[t]{0.9\textwidth}
   \begin{flushleft}
\setlength{\baselineskip}{12pt}
{\slshape  {\footnotesize{
$^{a}$Universit\`{a}  {\em {C\`{a} Foscari}}, Department of Molecular Sciences and Nanosystems, 
Via Torino 155/B, I-30172 Venezia, Italy, 
and \\
$^{b}$ Universit\`{a} di Padova, Dipartimento di Fisica {\em {G. Galilei}}, 
 Via Marzolo 8, I-35131 Padova, Italy.
}}}\\
 \footnotesize{$^*$salvino.ciccariello@unipd.it}
\end{flushleft}
\end{minipage}
}      
%
\date{\today\\ 
}

\maketitle                        
\begin{abstract} \noindent 
Using some rigorous results by Wiener [(1930). {\em  Acta Math.} {\bf 30}, 118-242] on the Fourier integral of a bounded function and the condition 
that small-angle scattering intensities of amorphous samples are almost 
everywhere continuous, we obtain the conditions that must be obeyed by 
a function  $\eta(\br)$ for this may be considered a physical scattering 
density fluctuation. It turns out that these conditions can be recast in 
the form that the $V\to\infty$ limit of the modulus of the Fourier transform  
of $\eta(\br)$, evaluated over a cubic box of volume $V$ and divided by $\sqrt{V}$, exists and that its square obeys the Porod invariant relation. 
Some examples of one-dimensional scattering density functions, obeying 
the aforesaid condition, are also numerically illustrated.\\

\noindent Synopsis: {\em  A function can be considered a physical 
scattering density fluctuation if  the modulus of 
its  Fourier transform, evaluated   over a finite volume $V$ and divided by $\sqrt{V}$, tends to a bounded non-vanishing function as 
$V\to\infty$ and if the squared limit obeys the Porod invariant relation.}\\

\end{abstract}
\vfill
\eject
{}{}
\subsection*{ 1. Introduction}  
In classical physics the scattering theory is based on the assumption that a sample is 
characterized by a scattering density function $n(\br)$ (Guinier, 1952; 
Guinier \& Fournet, 1955, Feigin \& Svergun, 1987), 
the existence of which is assumed without discussing its mathematical properties. 
The aim of this paper is to discuss this aspect. \\ 
Let us briefly recall the main equations that allow us to pass from the 
scattering density function to the scattering intensity. The main step is 
the fact that $dv$, the sample's infinitesimal volume element  set at 
the point $\br$, contributes to the scattering amplitude with $n(\br)e^{i\bq\cdot\br}dv$   [the modulus $q$ of the scattering 
vector $\bq$ being related to the scattering angle $\vartheta$ and 
the wavelength $\lambda$ of the ingoing beam radiation  by the 
relation  $q=(4\pi/\lambda)\sin(\vartheta/2)$]. The normalized 
scattering intensity $I_V(\bq)$ is the square modulus of the total 
scattering amplitude divided by volume of the illuminated portion 
of the sample and the intensity of the ingoing beam. Hence, 
setting  the last quantity equal to one, it is simply given 
by the expression
\begeq\label{1.1} 
I_V(\bq) = |\tn_V(\bq)|^2/V, 
\endeq 
where $\tn_V(\bq)$, the Fourier transform (FT) of $n_V(\br)$, is 
defined as   
\begeq\label{1.2}
\tn_V(\bq)\equiv \int_{R^3} n_V(\br)e^{i\bq\cdot \br}{\rm d}v=
\int_V n(\br)e^{i\bq\cdot \br}{\rm d}v.
\endeq
Here  $V$ denotes either  the sample's illuminated spatial set 
[generically having a right parallelepipedic shape]  or this  
set's volume, depending on the context, and $R^3$ the full 
three-dimensional Euclidean space.  Besides, $n_V({\br})$ is 
defined according to 
\begeq\label{1.3}
n_V({\br})\equiv\begin{cases}n(\br), & \text{if} \quad \br\in V;\\
0,&\text{elsewhere}.\end{cases}\endeq 
One experimentally finds that:\\ 
{\bA) -} once $V$ exceeds a few of $\rm{mm}^3$, $I_V(\bq)$ 
no longer depends on $V$, and\\  
{\bB) -} if the position of the sample is varied with respect to the ingoing beam, 
no change in the collected scattering intensity is observed.\\
To make the last point evident, one substitutes $V$ with  $V_O$ in the three above equations to emphasize the fact the illuminated part of the sample 
depends on the position of its gravity center $O$ even though its volume 
remains equal to $V$. Then,  on a mathematical ground, experimental results A and B imply  the existence of  $\lim_{V\to\infty}I_{V_O}(\bq)$ 
and that this value does not depend on $O$, \ie 
\begeq\label{1.4}
I(\bq)=\lim_{V\to\infty}I_{V_O}(\bq)= 
\lim_{V\to\infty}|\tn_{V_O}(\bq)|^2/V.
\endeq
Quantity $I(q)$ is commonly referred to as the normalized scattering 
intensity. This quantity is not directly observable  
because the counter pixels have a finite size (instead of a vanishing one, as 
we assumed above). 
This property implies that each pixel (say 
the  $i$th) subtends a solid angle $\Delta \Omega_i$.   
Consequently, the signal $\cI_i $ collected by the $i$th pixel is an 
integrated intensity given by 
\begeq\label{1.5}
{\cI}_i=\int_{\Delta \Omega_i} I(\bq(\hw))d\hw
\endeq 
with  $\bq(\hw)\equiv (2\pi/\lambda)(\hw-\hw_{in})$, where $\hw_{in}$ specifies the direction of the ingoing beam  and $\hw$  that of the  unit 
vector going from the sample (now treated as a point-like object since its largest diameter is much smaller than the distance of the sample from the 
counter) to a point of the pixel. It is commonly assumed that $I(\bq(\hw))$ 
be equal to $\cI_i/\Delta \Omega_i$ within the angular range 
$\Delta \Omega_i$.[Once more, depending on the context, 
$\Delta \Omega_i$ denotes the set or the size.] Then, the resulting  $I(\bq)$ has a histogram shape 
and  presents discontinuity points of the first order. One can 
compare set $\{\cI_i\}$ of the integrated intensity values 
collected with a counter to another set $\{{\cI_i}'\}$  collected either 
by the outset but slightly shifted counter, so as to have 
$\Delta \Omega_i'\ne \Delta \Omega_i$, or by a different counter 
with a better angular resolution, so as to have 
$\Delta \Omega_i'\subset \Delta \Omega_i$.  In both cases one finds that 
the discontinuity points of $I(\bq)$ and $I'(\bq)$ are different and this 
indicates that most of the discontinuities are not intrinsic to $I(\bq)$.  Moreover, 
the heights of the jumps show a different behavior depending on 
the analyzed material. In fact, for some materials one finds that the 
amplitudes are approximately the same in the first case (\ie\,  
$\Delta \Omega_i'\ne \Delta \Omega_i$) 
and reduced by about a $\Delta \Omega_i'/\Delta \Omega_i$ factor 
in the second. For other materials and mainly in the only second case, 
one observes that some jump heights remain 
approximately the same while the remaining ones are reduced by 
about the previous factor. This phenomenon is easily understood 
if $I(\bq)$ is written as 
\begeq\label{1.6}
I(\bq)=I_c(\bq) +\sum_j c_j\delta(\bq-\bq_j)
\endeq
where $I_c(\bq)$ is a function (with possible discontinuities of the 
only first kind),  $\delta(\cdot)$ denotes the Dirac function and 
the $c_j$s and the $\bq_j$s are a set of positive constants and 
a set of scattering vectors, respectively.  From  expression 
(\ref{1.6}), it follows that if, say,  the 
pixels $\Delta\Omega_i$ and $\Delta\Omega_k'$ contain
the  $\bq_j$ vector, then $\cI_i\approx\cI_k'\approx c_j$.  
This explains why the jumps, related to $\bq_j$, remain unaltered 
by the use of two different counters. The materials presenting the 
first kind of behavior, \ie\ with no Dirac-like contributions,  are 
amorphous while those presenting the second kind of behavior 
present a certain degree of crystallinity. In the following we shall 
confine ourselves to the theory of small angle scattering (SAS)  from 
amorphous samples. Thus, on the basis of the above experimental 
results, it will be assumed that:\\
{\bC) -} {\em the small angle scattering intensity, relevant to any amorphous 
sample,  is a function almost everywhere continuous throughout the 
full $q$ range.}\\  
It must be stressed that the SAS range of the observable $q$s is much 
more limited since it goes from $10^{-5}$ till $1$\AA$^{-1}$. If the last 
upper bound were increased by a  factor $10^2$ or more, the physics 
would drastically change because quantum and particle production 
effects can no longer be forgotten. In other words, the notion of 
scattering amplitude physically will no longer apply in the above 
reported form (see, \eg, Ciccariello, 2005). But, mathematically, the 
consideration of the FT for infinitely large values of $q$ does still make 
sense, and the validity of {\bC}  at large $\bq$s will only be considered 
from the last point of view. Concerning the reported lower bound, its 
presence is dictated by the necessity of introducing a beam stop in 
the experimental apparatus. The related limitation on the observable 
$q$s can to a large extent be overcome.  In fact, as discussed in detail by Guinier \& Fournet (1955) and, more recently, by Ciccariello (2017), 
in the experimentally accessible angular range the scattering intensity 
can be expressed in terms of the so called scattering density fluctuation 
$\eta(\br)$ that is simply obtained by subtracting to $n(\br)$ the mean 
value of the last quantity [\ie\, $\eta(\br)\equiv [n(\br)-\langle n\rangle]$, 
see equation (\ref{2.1}) below]. Once \bA\ is obeyed, the relation existing between $I_V(\bq)$ 
and $\eta(\br)$ is identical to equation (\ref{1.1}) in so far it reads
\begeq\label{1.7}
I_V(\bq)=|\teta_V(\bq)|^2/V
\endeq 
where $\teta_V(\bq)$ and $\eta_V(\br)$ are defined as $\tn_V(\bq)$ and 
$n_V(\br)$ in equations (\ref{1.2}) and (\ref{1.3}).  Adopting definition 
(\ref{1.1}) one would find that the right hand side (rhs) diverges as 
$\bq\to 0$. By contrast,   the $\bq\to 0$ limit of the rhs of (\ref{1.7})  
is finite since it is related to the isothermal compressibility of the 
analyzed sample. Then, hereinafter, the adopted definition of scattering 
intensity will be given by equation (\ref{1.7}) or, more precisely, 
by the $V\to\infty$ limit of (\ref{1.7}). Besides, the $I(\bq)$ values 
in the angular range behind the beam stop, by assumption, are obtained 
extrapolating toward the origin of reciprocal space the observed 
$I_V(\bq)$s. In this way, the sense of assumption \bC, for what concerns 
the full $\bq$ range, is fully clarified. 
Combining now assumption \bC\ with assumptions {\bA} and 
{\bB}, from (\ref{1.7}) one physically concludes that the FT $\teta_V(q)$ 
must be such that the following limit 
\begeq\label{1.8}
I(\bq)\equiv \lim_{V\to\infty}\Big|\frac {\teta_V(\bq)}{\sqrt{V}}\Big|^2
\endeq 
exists throughout the full $\bq$ range so as to define a function that 
is almost everywhere (a.e.) continuous function.  This relation implies that the 
scattering density fluctuation  $\eta(\br)$ relevant to any amorphous 
sample is such that the modulus of its FT, evaluated 
over a cubic set of volume V, must diverge as $V^{1/2}$ as $V\to\infty$,  because in this way only relation (\ref{1.8}) can hold true. 
Clearly, condition (\ref{1.8}) considerably restricts the class of functions eligible to be the scattering density fluctuation of an amorphous sample. 
For instance, any $\eta(\br)$  which is squared modulus integrable (\ie\ 
an $L_2$) function is  a trivial  scattering density fluctuation because it 
yields a vanishing scattering intensity,  since  its FT exists 
[see, \eg, Chandrasekharan, (1980)] and, consequently, once it is divided 
by $V^{1/2}$,  limit (\ref{1.8})  vanishes. Another  trivial one-dimensional 
example is the function $\eta(x)= \sin^2(x)/|x|^{1/2}$ [see eq. (40) 
of Ciccariello, 2017] that is not $L_2$. Its FT over 
the interval $[-L,\,L]$ is 
\begin{eqnarray}
&&\quad  -\sqrt{\frac{\pi}{4Lq(q^2-4)}}\Biggl[\sqrt{q(q+2)}\,\,
C\Bigl(\sqrt{\frac{2L(q-2)}{\pi}}\Bigr) -\nonumber \\
&&  2\sqrt{q^2-4}\,\,
C\Bigl(\sqrt{\frac{2Lq)}{\pi}}\Bigr) +\sqrt{q(q-2)}\,\,
C\Bigl(\sqrt{\frac{2L(q+2)}{\pi}}\Bigr)\Biggr]\nonumber
\end{eqnarray}
where $C(\cdot)$ denotes the Fresnel  cosine integral 
(see, \eg, section 8.25 of  Gradshteyn \& Ryzhik, 1980).  
Using the asymptotic expression of the last function one easily verifies 
that, in the $L\to\infty$ limit,  the above expression vanishes 
if $q\ne 0$ and is equal to 1 if $q=0$.  These examples and other 
unsuccesfull attempts made by the present authors indicate that 
it is not trivial to find $\eta(\br)$ examples that obey equation 
(\ref{1.8}).  The basic question is then: which are the properties of 
$\eta(\br)$ for conditions {\bf A-C} to be fulfilled?\\ 
Since scattering density fluctuations result from statistical mechanics 
averages (see, \eg, Morita \& Hiroike, 1961), they are  expected to be  bounded functions.  
[In this respect it is noted that  point-like (and, therefore, unbounded)  
scattering densities are often considered (Guinier, 1952) in dealing with 
crystalline materials. This  only is a  mathematical idealization, 
indeed quite useful   because all the spatial configurations 
of the scattering centers, compatible with a given scattering intensity, 
can, at least in principle,  be determined by the associated algebraic 
approach (see, \eg, Cervellino \& Ciccariello, 2001).]   
Besides, scattering density fluctuation are also expected to  contain 
some randomness elements owing to their statistical origin. This 
condition is quite hard to be translated into a mathematical 
definition but it suggests that the definition of scattering density 
fluctuations must conform to mathematical rules that are in some 
way probabilistic. \\ 
This paper will focus on these aspects. Using some rigorous 
mathematical results derived by Wiener in two papers (1930) and (1932), 
hereinafter referred to as I and II, it will be 
shown  that:\\
{\bSa -} {\em  any function $\eta(\br)$ can be considered 
a scattering density fluctuation if it has vanishing  mean value, 
its FT is such that the limit reported on the  rhs of 
equation (\ref{1.8})  exists  and  obeys to  }
\begeq\label{1.9}
\int_{R^3}I(\bq)dv_q=(2\pi)^3\langle \eta^2\rangle,
\endeq where $dv_q$ denotes the infinitesimal volume of 
reciprocal space and $\langle \eta^2\rangle$ the mean value of the 
squared density fluctuation. The above relation is known as 
Porod's (1951) invariant relation.\\ 
The plan of the paper, that, for simplicity, is confined to 
one dimensional samples, is as follows. 
Section 2 deals with the introduction of the scattering density 
fluctuation $\eta(x)$ and the definition and some general 
properties of the associated correlation functions $\gamma(x)$ 
and $\gamma_L(x)$, which respectively refer 
to the infinitely large sample and to the sample of size $2L$. 
Section 3 essentially reports the basic work of Wiener (1930) who 
showed how to rigorously define the integrated scattering intensity 
starting from a generalized Fourier integral of $\gamma(x)$. 
We show that it is possible to achieve a more detailed characterization 
of functions $\eta(\br)$ using assumption {\bC}. In particular, it 
turns out that the $\gamma(x)$ of any amorphous sample  is a 
continuous and an $L_2$ summable function.  Section 4 shows the 
equivalence of this analysis and statement \S1. Moreover, it also 
reports two examples of scattering density fluctuations of the 
M\'ering-Tchoubar (1968) kind that obey relation (\ref{1.8}). Finally, 
section 5 draws  the final conclusions. 
\subsection*{2. Definition and properties of the correlation function}
The scattering density fluctuation (SDF) $\eta(x)$, associated to a 
scattering density function $n(x)$, is simply obtained by subtrating 
to $n(x)$ the mean value $\mn$ of $n(x)$ evaluated all over the 
space. Thus, 
\begeq\label{2.1}
\eta(x) \equiv n(x)-\mn
\endeq
with 
\begeq\label{2.2} 
\mn \equiv \lim_{L\to\infty}\frac{1}{2L}\int_{-L}^L n(x) dx=  
\lim_{L\to\infty}\frac{1}{2L}\int_{-\infty}^{\infty} n_L(x) dx,
\endeq 
where, similarly to definition (\ref{1.3}), $n_L(x)$ is equal to 
$n(x)$ if $|x|\le L$ and to zero elsewhere.  The $\mn$ value generally 
is finite and different form zero.   
From (\ref{2.1}) and (\ref{2.2}) follows that $\langle \eta\rangle$, the 
mean value of $\eta(x)$, vanishes. As already anticipated, 
it is physically  by no way  restrictive to assume that:\\   
\ba) $|\eta(x)|$ is a bounded function (\ie\ $|\eta(x)|\le B$ whatever $x$ with  
$B>0$), integrable in the Lebesgue sense, and \\
\bb) it  has no limit as $x\to \pm\infty $ so that  $\eta(x)$ is 
expected to irregularly oscillate around zero.\\  
One should note that the last condition, on the one hand, excludes that 
$\eta(x)$ may be an $L_2$ function and, on the other hand, it introduces 
some sort of randomness through the irregularity of the oscillations.  
 Assumption \ba) ensures the validity of the following property:\\
{\bf {P.1 -}} {\em {the mean value definition is translational invariant.}} \\
In fact, the mean value evaluated over the  interval $[-L,\,L]$, translated 
by $a$, reads  
\begeq\label{2.3a}
\frac{1}{2L}\int_{-L+a}^{L+a} \eta(x) dx=\frac{1}{2L}\int_{-L-a}^{L+a} 
\eta(x) dx-\frac{1}{2L}\int_{-L-a}^{-L+a} \eta(x) dx. 
\endeq
Observing  that  
\begeq\label{2.3b}
\frac{1}{2L}\int_{-L-a}^{L+a} \eta(x) dx= 
\frac{2(L+a)}{2L}\frac{1}{2(L+a)}\int_{-L-a}^{L+a} \eta(x) dx
\endeq
and that 
\begeq\label{2.3c}
\Bigl|\frac{1}{2L}\int_{-L-a}^{-L+a} \eta(x) dx\Big|\le \frac{1}{2L}
\int_{-L-a}^{-L+a} |\eta(x)| dx \le \frac{a B}{L},
\endeq 
one concludes that, whatever $a$,  
\begeq\label{2.3d}
\lim_{L\to\infty}\Bigl[\frac{1}{2L}\int_{-L-a}^{L+a} \eta(x) dx\Bigr]=
\lim_{L\to\infty}\Bigl[\frac{1}{2L}\int_{-L}^{L} \eta(x) dx\Bigr]=\meta=0.
\endeq 
This result mathematically formulates the physical property that the 
mean value of a quantity, relevant to a macroscopically homogeneous 
material, is independent of the sampling, provided the sampled volume 
be  sufficiently large. \\ 
The correlation function of a bounded sample, with SDF $\eta(x)$ and 
size $2L$, is defined as 
\begeq\label{2.4}
\gamma_L(x)\equiv \frac{1}{2L}\int_{-\infty}^{\infty}\eta_L(x+y){\overline{\eta_L}}(y)dy
= \frac{1}{2L}\int_{-L}^{L}\eta_L(x+y){\overline{\eta_L}}(y)dy.
\endeq 
where the overbar denotes the complex conjugate and $\eta_L(\cdot)$ 
is function $\eta(\cdot)$ restricted to the interval $[-L,\,L]$. This 
definition, due to Wiener (1930),  generalizes the one adopted in 
SAS theory generally dealing with real $\eta(x)$s. It is also noted that 
one migth adopt the alternative definition 
\begeq\label{2.4b}
\Gamma_L(x)\equiv  \frac{1}{2L}\int_{-L}^{L}\eta(x+y){\overline{\eta}}(y)dy.
\endeq 
 If the SDF obeys condition \ba),   the two definitions 
coincide in the limit $L\to \infty$, \ie 
\begeq\label{2.4c}
\lim_{L\to\infty}\Gamma_L(x) = \lim_{L\to\infty}\gamma_L(x).  
\endeq 
In fact, one finds that 
\begeq\label{2.4d}
\Gamma_L(x)-\gamma_L(x)= \frac{1}{2L}\int_{L-x}^{L}\eta(x+y){\overline{\eta(y)}}dy, \quad {\rm if}\quad x>0, 
\endeq 
and, since 
\begeq\label{2.4e}
\Big| \frac{1}{2L}\int_{L-x}^{L}\eta(x+y){\overline{\eta(y)}}dy\Big|\le 
 \frac{1}{2L}\int_{L-x}^{L}B^2dy=\frac{x}{2L}, 
\endeq
one immediately realizes that relation (\ref{2.4c}) is true because the case $x<0$ can similarly be handled. For brevity, the functions $\gamma_L(x)$ 
and $\Gamma_L(x)$ will be named limited correlation functions (LCF) to distinguish them from the correlation function (CF) relevant to the infinitely large sample.  This  is defined according to 
\begeq\label{2.5}
\gamma(x)=\lim_{L\to\infty}\gamma_L(x)=\lim_{L\to\infty}\Gamma_L(x) .
\endeq  
It will hereinafter be assumed that $\gamma(x)$ exists for any real $x$ value. It is also true the property that:\\    
{\bf {P.2 -}} {\em{ 
The definition of LCF becomes translational invariant in the $L\to\infty$ 
limit. }}\\
In fact, if $a>0$, the difference 
between the translated and the outset definition of $\Gamma_L$ yields   
\begin{eqnarray} 
&&\Big|\frac{1}{2L}\int_{-L+a}^{L+a}\eta(x+y){\overline{\eta}}(y)dy- \frac{1}{2L}\int_{-L}^{L}\eta(x+y){\overline{\eta}}(y)dy\Big|=\nonumber\\
&&\frac{1}{2L}\Big|\int_{L}^{L+a}\eta(x+y){\overline{\eta}}(y)dy-\int_{-L}^{-L+a}\eta(x+y){\overline{\eta}}(y)dy\Big|\le \nonumber\\
&&\frac{1}{2L}\Bigl[\int_{L}^{L+a}B^2dy+\int_{-L}^{-L+a}B^2dy\Bigr]=\frac{aB^2}{L}.  \nonumber
\end{eqnarray} 
The rightmost value tends to zero as $L\to\infty$ and the invariance is 
proved because the discussion of the case $a<0$ is quite similar. \\ 
The assumed boundedness of the SDF ensures that  $\gamma_L(x)$ exists, 
is continuous everywhere and only differs from zero within the interval 
$[-2L,\, 2L]$.  However, the continuity property does not generally apply 
to $\gamma(x)$. 
An example, due to Wiener (II, pag. 151), makes this point evident.  Assume that 
$\eta(x)=e^{ix^2}$ so as to obey condition \ba). One finds that 
\begeq\nonumber 
\Gamma_L(x)=\frac{1}{2L}\int_{-L}^{L}e^{i[(x+y)^2-y^2]}dy=
\frac{e^{ix^2}}{2L}\int_{-L}^{L}e^{2i x y}dy
\endeq
The last integral is equal to $2L$ if $x=0$ and to $\sin(2xL)/x$ if $x\ne 0$. 
Hence, in the $L\to\infty$ limit, one gets: $\gamma(x)=1$ if $x=0$ and  
$\gamma(x)=0$ if $x\ne 0$ and the CF is not continuous. This example 
shows that if one requires that $\gamma(x)$ exists,  the class of the 
$\eta(x)$ functions must obey appropriate mathematical constraints. 
In fact, the condition that $\eta(x)$  is Lebesgue summable over any 
compact domain and everywhere bounded only is a sufficient condition.  
It is not easy to work out these constraints. They are implicitly defined assuming that:\\   
{\bf{P.3 - }} {\em the  $\eta(x)$ functions that can be SDFs 
belong to the set of functions that  yield CFs that are defined 
throughout $(-\infty,\,\infty)$.} \\ 
This set of functions will be denoted by $\cW$ (in honor of 
Wiener who instead used symbol $S$.)\footnote{To make the reference to 
papers I and II easier, we note that Wiener's most frequently used 
symbols have been converted to ours according to: 
$f(x)\to\eta(x)$, $\phi(x)\to\gamma(x)$, $S\to\cW$, $S'\to\cW'$. Two 
further sets of functions $\cWb$ and $\cWp$ will later be introduced. 
The sets are related among themselves according to $\cWp\subset \cWb\subset\cW$ and $\cWp\subset \cW'\subset\cW$}.  
We note now that:\\    
{\bf{P.4 - }} {\em {The  function set  $\cW$ is a 
vectorial space.}}\\
To  prove this property  one has to show that: 
1)  if $\eta(x)\in\cW$ then $\alpha\eta(x)\in \cW$ for any  complex 
number $\alpha$, and 2) if $\eta_1(x)$ and $\eta_2(x)\in\cW$ 
then $[\eta_1(x)+\eta_2(x)]\in \cW$. The first condition is obviously true. 
The proof of the second will slightly be postponed after we have reported 
two important theorems of Wiener (see II, pag. 154-156). The first states 
an important property, given for granted without any proof in SAS textbooks (Guinier \& Fournet, 1955; Feigin \& Svergun, 1987), \ie:\\    
 {\bf{P.5 - }} {\em if $\eta(x)\in\cW$, the correlation CF whatever $x$ obeys the inequality}
\begeq\label{2.6}
|\gamma(x)|\le \gamma(0)=\langle |\eta|^2\rangle, 
\endeq 
where the rightmost equality follows from defintion (\ref{2.5}) since  
$\langle |\eta|^2\rangle$ denotes the mean of the squared SDF. 
The second that: \\
 {\bf{P.6 - }} {\em the correlation CF is everywhere continuous  
if it is continuous at $x=0$.}\\
The proof of {\bf{P.5}} is somewhat simpler than that of Wiener (II, 
pag. 154) if, in agreement with \ba), one confines himself  to $\cWb$, 
\ie\, the subset of the  bounded functions belonging to $\cW$. In fact,  
from definition (\ref{2.4}), by the Schwarz inequality one gets 
\begin{eqnarray}\label{2.7}
|\gamma_L(x)|^2&=&\Big|\frac{1}{2L}\int_{-L}^{L}\eta_L(x+y)
{\overline{\eta_L(y)}}dy\Big|^2\le \\
\quad &&\Big[\frac{1}{2L}\int_{-L}^{L}|\eta_L(x+y)|^2dy\Big]
\Big[\frac{1}{2L}\int_{-L}^{L}|\eta_L(y)|^2dy\Big].
\end{eqnarray}
One also has
\begin{eqnarray}\label{2.8}
&&\int_{-L}^{L}|\eta_L^2(x+y)|dy\le \int_{-L}^{L}|\eta^2(x+y)|dy=
\int_{-L+x}^{L+x}|\eta^2(t)|dt=\nonumber \\
&&\int_{-L-x}^{L+x}|\eta^2(t)|dt-\int_{-L-x}^{-L+x}|\eta^2(t)|dt\le
 \int_{-L-x}^{L+x}|\eta^2(t)|dt+\int_{-L-x}^{-L+x}|\eta^2(t)|dt\le \nonumber\\
&& \quad\quad\quad \int_{-L-x}^{L+x}|\eta^2(t)|dt+2x B^2,
\end{eqnarray}
where, in obtaining the last relation, we used the fact that $|\eta(x)|\le B$.  
After substituting the above inequality in equation (\ref{2.7}) one gets 
\begin{eqnarray}\label{2.9}
&&|\gamma_L(x)|^2\le \Big[\frac{1}{2L}\int_{-L}^{L}|\eta_L^2(y)|dy\Big]\times\nonumber\\
&&\Big\{\frac{2(L+x)}{2L}\Big[\frac{1}{2(L+x)}\int_{-(L+x)}^{L+x}|\eta^2(y)|dy\Big]+\frac{xB^2}{L}\Big\}.
\end{eqnarray}
In the $L\to\infty$ limit, the quantities inside the square brackets approach 
to $\gamma(0)$ while  $xB^2/L$ and $2(L+x)/2L$ tend to zero and one, 
respectively. In this way  {\bf{P.5}} is proved.  
To prove {\bf{P.6}}, one starts from 
\begeq\label{2.10}
|\gamma(x+\epsilon)-\gamma(x)|^2=\lim_{L\to\infty}\Big|\frac{1}{2L}
\int_{-L}^{L}[\eta(x+\epsilon+t)-\eta(x+t)]{\overline\eta}(t)dt
\Big|^2.
\endeq
The above rhs  (leaving momentarily aside the limit),  by  Schwartz' 
inequality, does not exceed the quantity 
\begeq\label{2.11}
\Bigl[\frac{1}{2L}\int_{-L}^{L}|\eta(x+\epsilon+t)-\eta(x+t)|^2dt\Bigr]\Bigl[\frac{1}{2L}
\int_{-L}^{L}|\eta|^2(t)dt\Bigr]. 
\endeq
Expanding the integrand of the first integral, one gets  
four addends: $|\eta(x+\epsilon+t)|^2$, $|\eta(x+t)|^2$, 
$-\eta(x+\epsilon+t){\overline{\eta(x+t)}}$ and $-{\overline{\eta(x+\epsilon+t)}}\eta(x+t)$. 
The integration of each of these terms over $[-L,\,L]$, the subsequent limit $L\to\infty$ and the established translational invariance property 
respectively yield: $\gamma(0)$, $\gamma(0)$, $-\gamma(\epsilon)$ and $-{\overline{\gamma(\epsilon)}}$. 
Substituting these findings in the rhs of  (\ref{2.10}) one obtains 
\begeq\label{2.12}
|\gamma(x+\epsilon)-\gamma(x)|^2\le 
\bigl[2\gamma(0)-\gamma(\epsilon) -{\overline{\gamma(\epsilon)}}\bigr]\gamma(0), 
\endeq
which proves the theorem. \\ 
We complete now the proof that $\cWb$ is a vectorial space by showing  
that condition 2 also is obeyed. To this aim one must show that the $L\to\infty$ limit of 
\begeq\label{2.13}
\frac{1}{2L}\int_{-L}^{L}[\eta_1(x+t)+\eta_2(x+t)]\overline{[[\eta_1(t)+\eta_2(t)]}dt
\endeq 
exists if $\eta_1$ and $\eta_2$ belong to $\cWb$.  As it was done in 
equation (\ref{2.11}), expanding the integrand one finds the LCFs   $\Gamma_{1;L}(x)$ and $\Gamma_{2;L}(x)$, relevant to the SDFs 
$\eta_1(x)$ and $\eta_2(x)$, the integral 
\begeq\label{2.14}
\Gamma_{1,2;L}(x)\equiv \frac{1}{2L}
\int_{-L}^{L}\eta_1(x+t)\overline{\eta_2(t)}dt
\endeq
and the integral $\Gamma_{2,1;L}(x)$. Since $\eta_1$ and $\eta_2$ 
belong to $\cWb$, the $L\to\infty$ limits of  $\Gamma_{1,L}(x)$ 
and $\Gamma_{2,L}(x)$ will yield the 
CFs $\gamma_{1}(x)$ and  $\gamma_{2}(x)$. 
By Schwartz' inequality and the procedure followed to prove {\bf P.2},  
in the $L\to\infty$ limit  one gets 
\begeq\label{2.15}
|\Gamma_{1,2;L}(x)|\to |\gamma_{1,2}(x)| \le \sqrt{\gamma_1(0)\,\gamma_2(0)}
\endeq
and an identical relation for $|\Gamma_{2,1;L}(x) |$. In this way the 
proof of condition 2 is accomplished. Thus, $\cWb$ is a vectorial 
space and in the same way one proves that $\cW$ also is a vectorial 
space. \\ 
We report some examples of SDFs which 
yield algebraically known CFs (II, pag. 151):\\
1 - if $\eta(x)\in L_2(-\infty,\infty)$, the associated CF identically 
vanishes. In fact, from the hypothesis follows that 
$\lim_{L\to\infty}\int_{-L}^{L}|\eta(x)|^2dx=
\int_{-\infty}^{\infty}|\eta(x)|^2dx<\infty$. Then,   $\gamma(0)=0$ 
by (\ref{2.5}) and $\gamma(x)\equiv 0$ by (\ref{2.6});\\
2 - with  the choice $\eta(x)=e^{i\alpha x}$ (with $\alpha\in R$), 
which is a bounded function with  mean value equal to zero 
if $\alpha\ne 0$ and to one if $\alpha=0$,  by equations (\ref{2.5}) and 
(\ref{2.4b}) one finds that the associated CF again is  $e^{i\alpha x}$ 
which is continuous throughout $R$;\\
3 - with the choice $\eta(x)=e^{i\alpha|x|^{a}}$ (with $\alpha$ real 
and $0<a<1$), generalizing  the method reported at page 151 of II,  
the resulting CF turns out to be identically equal to 1;\\
4 - we already saw that the choice $\eta(x)=e^{i\alpha x^2}$ (with 
$\alpha$ real) yields: $\gamma(x)=1$ if $x=0$ and $\gamma(x)=0$ if 
$x\ne0$, and the CF is now discontinuous by contrast to the previous 
cases. \\
Examples 1, 3 and  4 show that the problem: 
{\em "which is the function $\eta(x)$ which yields an assigned CF?  
or, equivalently, which is the $\eta(x)$ solution of the non linear 
integral equation} \\
\begeq\label{2.15a}
\lim_{L\to\infty}\frac{1}{2L}\int_{-L}^{L}\eta(x+y){\overline{\eta(y)}}dy=\gamma(x)
\endeq
has an infinity  of solutions.  This conclusion is further 
strengthned by the property:\\ 
{\bf{ P.7 -}} {\em{if $\eta(x)$ is a solution of equation (\ref{2.15a}),  
any function of the kind  $[\eta(x+a)+\eta_1(x)]$, with $a\in R$ and  
$\eta_1(x)\in L_2(-\infty,\,\infty)$,  is solution of the 
same equation.}}\\ 
In fact, the proof that, for any real $a$,  $\eta(x+a)$ reproduces the same 
CF of $\eta(x)$ is a consequence of {\bf P.2}. If 
$\eta_1(x)\in L_2(-\infty,\,\infty)$, the proof that the function 
$\eta(x)+\eta_1(x)$ has the same CF of $\eta(x)$  immediately 
follows from example 1 and inequality (\ref{2.15}). Finally, the 
linearity of $\cW$ ensures that the property holds also true 
for $[\eta(x+a)+\eta_1(x)]$.
\subsection*{3. Wiener's results on   the SDF and  CF  Fourier integrals}
We describe now the procedure followed by Wiener to rigorously 
evaluate the integrated scattering intensity. Some of his results, 
once combined with assumption \bC,  are quite important for the 
theory of SAS from amorphous 
samples. Having confined ourselves to one-dimensional 
samples,  equation (\ref{1.4}) takes now the form 
\begeq\label{3.1a}
I(q)=\lim_{L\to\infty}I_L(q)=\lim_{L\to\infty}|\teta_L(q)|^2/2L,
\endeq 
where $\teta_L(q)$ denotes  the FT of the SDF $\eta(x)$ evaluated 
over the interval $[-L,\,L]$.   According to assumption \ba),  $\eta_L(x)$,   
having the interval $[-L,\,L]$ as support, is a bounded  
 $L_1(-L,\,L)$ function. 
Then, its Fourier transform 
\begeq\label{3.1}
\teta_L(q)\equiv\int_{-\infty}^{\infty}\eta_L(x)e^{iqx}dx
\endeq 
exists and is continuous for any $q$ value [see, \eg, Chandrasekharan 
(1989), pag. 2].  
However, the limit $\lim_{L\to\infty}\teta_L(q)$ does not generally 
exist, because  $\eta(x)$ does not approach zero at $x=\pm\infty$ and, therefore,  it  belongs neither to $L_1(-\infty,\,\infty)$ nor
 to $L_2(-\infty,\,\infty)$.  The non existence of 
 $\lim_{L\to\infty}\teta_L(q)$ is essential for equation 
(\ref{3.1a}) to make sense otherwise, if it would exist, one 
would find a vanishing scattering intensity.\\ 
This difficulty also applies to the FT of the CF generated by $\eta(x)$. 
In fact, equation(\ref {3.1a}) can be written, by definition (\ref{2.4}), 
as 
\begin{eqnarray}\label{3.2}
&& I(q)=\lim_{L\to\infty}\frac{1}{2L}\int_{-\infty}^{\infty}
\int_{-\infty}^{\infty}\eta_L(x)
{\overline{\eta_L(y)}}e^{iq(x-y)}dxdy=\nonumber \\
&&\quad\quad\quad \lim_{L\to\infty}\int_{-\infty}^{\infty}e^{iqt}
\gamma_L(t). 
\end{eqnarray}
According to {\bf{P.5}}, so far it is only known that $\gamma(x)$ is bounded 
if $\eta(x)\in\cWb$, and this condition is not sufficient to ensure the 
existence of the rightmost limit present in (\ref{3.2}).  
Wiener (I, pag. 134)  overcame the difficulty of the non 
existence of the above limit as well as of the $L\to\infty$ limit of 
$\teta_L(q)$ (I, pag. 151) introducing a factor of convergence within the  
relevant integrals. In fact, instead of $\teta_L(q)$, Wiener (II, pag. 138)  considered function $s_{L}(q)$  defined as  
\begin{eqnarray}\label{3.3a}
s_L(q)&\equiv&s_{L,1}(q)+s_2(q)\label{3.3a}\end{eqnarray} 
with 
\begin{eqnarray}\label{3.3}
s_{1,L}(q)&\equiv&\int_{1}^{\infty}\eta_L(x)\frac{e^{iqx}}{ix}dx+
\int_{-\infty}^{-1}\eta_L(x)\frac{e^{iqx}}{ix}dx,\\
s_2(q)&\equiv& \int_{-1}^{1}\eta(x)\frac{e^{iqx}-1}{ix}dx.\label{3.4}
\end{eqnarray} 
[Omitting the factor $(2\pi)^{-1/2}$ we have slightly changed  Wiener's definition.] In (\ref{3.3}) the integration bounds $\pm\infty$ 
can be set equal to $\pm L$. As far as $L$ is finite, functions 
$s_{L,1}(q)$ and 
$s_2(q)$ are entire functions of $q$ in the whole complex $q$ plane 
and one has 
\begeq\label{3.3b}
s'_{L}(q)=\teta_L(q),
\endeq
where the prime denotes the derivative. It is noted that equations 
(\ref{3.3}) and (\ref{3.4}) can be combined into the form of a FT, \ie 
\begeq\label{3.3d}
s_L(q+\epsilon)-s_L(q-\epsilon)=\int_{-\infty}^{\infty}\eta_L(x)
\frac{2\sin(\epsilon x)}{x}e^{i q x}dx,
\endeq
$\epsilon$ being a real number. Thus, $[s_L(q+\epsilon)-s_L(q-\epsilon)]$ 
is the FT of the $L_1[-L,\,L]$ function $2\eta_L(x){2\sin(\epsilon x)}/{x}$.
The above equation can be inverted to get
\begeq\label{3.3e}
\eta_L(x) =\frac{x}{4\pi\sin(\epsilon x)}\int_{-\infty}^{\infty}
[s_L(q+\epsilon)-s_L(q-\epsilon)]e^{-iqx}dq.
\endeq            
The  $L\to\infty$ limit is the crucial point because 
the corresponding limit of $\teta_L(q)$ generally does not exist since  
$\eta(x)$  neither belongs to $L_1$ nor to $L_2$. Actually, using the boundedness of $\eta(x)$ [\ie\, $|\eta(x)|\le B$],  from equation (\ref{3.3d}) 
one would get by theorem 1.5 of Chandrasekhar (1980) that the least 
upper bound, with respect to $L$,  of $[s_L(q+\epsilon)-s_L(q-\epsilon)]$ cannot exceed $4B\log L$ 
(a bound diverging with $L$), as $q$ ranges over $R$.  In reality, 
the last function is bounded and decreases at large $q$s. 
In fact,  the presence of the factor $x$ in the integrands' 
denominators of  (\ref{3.3})  ensures that  $\eta_L(x)/x$ 
belongs to  $L_2$ also in the $L\to\infty$ limit.  
Consequently, the integrals present in (\ref{3.3}) exist and converge 
in the mean (a condition specified hereinafter by the symbol 
"$\limM$") and, since integral (\ref{3.4}) converges uniformly,   
one can set 
\begeq\label{3.3c}
s_1(q)\equiv \limM_{L\to\infty} \,s_{1,L}(q)\quad{\rm {and}}
\quad s(q)\equiv \limM_{L\to\infty}\, s_L(q)=s_1(q)+s_2(q).
\endeq
From the last relation follows that, whatever $\epsilon(\in R)$, 
\begin{eqnarray}\label{3.5}
s(q+\epsilon)-s(q-\epsilon)&=&
\limM_{L\to\infty}[s_L(q+\epsilon)-s_l(q-\epsilon)\bigr]=
\nonumber\\
\quad && \limM_{L\to\infty}
\int_{-L}^{L}\eta_L(x)\frac{2\sin(\epsilon x)}{x}e^{iqx}dx, 
\end{eqnarray} 
because $\eta(x)\frac{2\sin(\epsilon x)}{x}$ is an $L_2[-\infty,\,\infty]$ function. Then, 
(\ref{3.5}) can be inverted to get the $\limM$ of equation (\ref{3.3e}), \ie
\begeq\label{3.5b}
\eta(x) =\frac{x}{4\pi\sin(\epsilon x)}\limM\int_{-\infty}^{\infty}
[s(q+\epsilon)-s(q-\epsilon)]e^{-iqx}dq,
\endeq  
and  by  the Plancherel theorem one also gets 
\begeq\label{3.5a}
\int_{-\infty}^{\infty}|s(q+\epsilon)-s(q-\epsilon)|^2dq=
8\pi \int_{-\infty}^{\infty}|\eta(x)|^2\frac{\sin^2(\epsilon x)}{x^2}dx. 
\endeq 
We recall now the Tauberian theorem (II, pag. 139) that states that:\\
{\bf {P.8 - }}{\em {if $F(x)$ is a non-negative 
function and if  one of the  two limits}} 
\begeq\label{3.51a}
\lim_{L\to\infty}\frac{1}{2L}\int_{-L}^{L}F(x)dx,\quad\quad 
\lim_{\epsilon\to 0}\frac{1}{\pi\epsilon}\int_{-\infty}^{\infty}F(x)\frac{\sin^2(\epsilon x)}{x^2}dx
\endeq
{\em {exists, then the other also exists and the two limits are equal.}} \\ 
The application of this theorem to equation (\ref{3.5a}) yields [see 
also (\ref{2.6})]
\begeq\label{3.51b}
\lim_{\epsilon\to\infty}\frac{1}{\epsilon}\int_{-\infty}^{\infty}
|s(q+\epsilon)-s(q-\epsilon)|^2dq=8\pi^2 \langle|\eta|^2\rangle=
8\pi^2 \gamma(0). 
\endeq
The case of the CF was handled with by Wiener (II, pag. 161) in a 
similar way. In fact, the generalized Fourier integral 
transform of $\gamma(r)$  reads
\begeq\label{3.6}
\sigma(q)=\limM_{L\to\infty}\Big\{
\Bigl[\int_{1}^{L}+\int_{-L}^{-1}\Bigr]\gamma_L(x)\frac{e^{iqx}}{ix}dx\Big\}+
\int_{-1}^{1}\gamma_L(x)\frac{e^{iqx}-1}{ix}dx. \quad
\endeq
Similarly to equations (\ref{3.5}), (\ref{3.5b}) and (\ref{3.5a}), one finds that  
\begin{eqnarray}\label{3.7}
\sigma(q+\epsilon)-\sigma(q-\epsilon)&=&
 \limM_{L\to\infty}\Bigl\{
\int_{-L}^{L}\gamma_L(x)\frac{2\sin(\epsilon x)}{x}e^{iqx}dx\Bigr\},   
\end{eqnarray}
\begeq\label{3.7b}
\gamma(x)=\frac{x}{4\pi \sin{\epsilon x}}\limM\int_{-\infty}^{\infty}
[\sigma(q+\epsilon)-\sigma(q-\epsilon)]e^{-i q x}dq,
\endeq
and
\begeq\label{3.7a} 
\int_{-\infty}^{\infty}|\sigma(q+\epsilon)-\sigma(q-\epsilon)|^2dq=
8\pi \int_{-\infty}^{\infty}|\gamma(x)|^2\frac{\sin^2(\epsilon x)}{x^2}dx. 
\endeq 
Wiener (see II, pag. 162) showed that $\sigma(q)$ and $s(q)$ are related as follows
\begeq\label{3.10}
\sigma(q)-\sigma(-q)={\rm{l.i.m.}}_{\epsilon\to 0}\Bigl\{\frac{1}{4\epsilon \pi}
\int_{-q}^{q}\big|s(u+\epsilon)-s(u-\epsilon) \big|^2du\Bigr\}. 
\endeq 
From this relation follows that $[\sigma(q)-\sigma(-q)]>0$ whatever $q$. 
Modifying the $\sigma(q)$ definition at a discrete set of points, the new $\sigma(q)$ (see I, pag. 136) can be chosen in such a way that it is a non-negative and  non-decreasing function of $q$. From now on we always refer  to this new $\sigma(q)$. From equations (\ref{3.10}) and (\ref{3.51b}), according to theorem 31 of Wiener [II, pag. 181],  it follows that:\\
{\bf{ P.9 - }} {\em  if $\eta(x)\in\cW$, then } 
\begeq\label{3.11b}
\sigma(\infty)-\sigma(-\infty)\le 2\pi \gamma(0)=
2\pi\langle |\eta|^2\rangle,
\endeq 
{\em and this inequality becomes an equality, \ie}
\begeq\label{3.11}
\sigma(\infty)-\sigma(-\infty)=2\pi \gamma(0),  
\endeq
{\em if $\eta(x)\in\cW'$, where $\cW'$ is the subset of $\cW$ formed 
by the functions that generate CFs that are everywhere continuous.} \\
The theorem that allows us to relate $\sigma(q)$ to the observed 
scattering intensity $I(q)$ is theorem 36 reported at the bottom of 
page 183 of II. It states:\\ 
{\bf {P.10 -}}  {\em {If  $\eta(x)\in\cW$,   the integral}}
\begeq\label{3.8}
S(q)=\int_{-\infty}^{\infty}\gamma(x)\frac{e^{iqx}-1}{i x}dx, 
\endeq 
{\em{exists almost everywhere and  one has a.e.}}
\begeq\label{3.9}
S(q)=\sigma(q)+const.
\endeq
Taking formally the $q$ derivative of $S(q)$ and recalling equation 
(\ref{3.2}) one finds that 
\begeq\label{3.9b}
S'(q)=\sigma'(q)=I(q)\ge 0
\endeq 
which in turns implies that $S(q)$ is the integrated scattered intensity 
or, adopting Wiener's terminology, $I(q)$ is the spectral density.  
Relation (\ref{3.9b}) does not make sense at the $q$ values where 
$\sigma(q)$ and, consequently, $S(q)$ jump. Nonetheless, one can still consider relation(\ref{3.9b})  true if one agrees to say that, at these 
$q$ values, $S'(q)$ and $I(q)$ behave as Dirac functions.  In the 
introductory section it was stated that SAS intensities originating from amorphous samples  do not have $\delta$-like contributions since, 
as stated in \bC, the observed $I(q)$s are a.e. 
continuous.  The request that assumption \bC\ be obeyed allows 
us to get a stricter characterization of  functions $\eta$s for these 
may be considered physical SDFs. In fact, the a.e. continuity of 
physical $I(q)$s, by equations (\ref{3.9b}) and (\ref{3.9}), implies that 
$\sigma(q)$ must be continuous and almost everywhere endowed of 
a derivative that is a.e. continuous. We recall now a theorem by Lebesgue 
[Kolmogorov \& Fomin (1980), pag. 340] which states that the derivative 
$f(x)=F'(x)$ of a  function $F(x)$ absolutely continuous in the interval 
$[a,\,b]$ exists, is summable and for any $x\in[a,\,b]$ it results $\int_a^xf(t)dt=F(x)-F(a)$.  Then, one must require the absolute continuity 
of the non-decreasing $\sigma(q)$ throughout $(-\infty,\,\infty)$ for the existence of the non-negative $\sigma'(q)$  
to be  ensured a.e. throughout $(-\infty,\,\infty)$. 
The last condition of absolute continuity in turns requires that physical 
SDF $\eta(x)$s are restricted to a subset of $\{\cWb\cap\cW'\}$ in such 
a way that the resulting $\gamma(x)$s and associated $\sigma(q)$s 
respectively are continuous and absolutely  continuous throughout the 
corresponding definition domains. 
This subset of $\{\cWb\cap\cW'\}$ will hereinafter be denoted 
by $\cWp$, where subscript {\em p} underlines that we are now dealing 
with  physical SDFs. Hence, the basic statement  (which was not reported by Wiener): \\ 
{{\bSb} -}  {\em {$\cWp$ is the subset of $\{\cWb\cap\cW'\}$ 
formed by the $\eta(x)$s that belong both to $\cWb$ and to $\cW'$ 
and that generate continuous CFs  such that the associated $\sigma(q)$s, defined by equation (\ref{3.6}),  are  non-decreasing and absolutely 
continuous throughout $(-\infty,\,\infty)$. Any  $\eta(x)\in \cWp$  
can be the physical SDF of an amorphous sample.}} \\ 
In the remaining part of this section we shall combine this statement 
with other results by Wiener  and in this way we shall show that: i) the 
scattering intensity and the CF of any amorphous sample are $L_2$ functions 
and each of them simply is the  FT transform of the other (see {\bf P.16}), 
and ii) any $\eta(x)\in \cWp$ contains no periodic contribution with a 
finite amplitude (see {\bf P.21}). \\ 
We begin by recalling equations (\ref{3.9}) and (\ref{3.9b}). One 
immediately realizes that:\\   
{\bf{P.11 -}} {\em{ 
equation (\ref{3.11}) is nothing else than the so-called Porod invariant relation since it can be recast in the form $\int_{-\infty}^{\infty}I(q)dq=
2\pi\langle|\eta|^2\rangle$.}}\\
This was, therefore, discovered by Wiener [see I, equation (4.10)] 
much earlier than Porod (1951). 
Another interesting result is theorem 32 of Wiener (II, pag. 181). It 
states that:\\   
{\bf{P.12 - }} {\em {if $\eta(x)\in\cW$, then}}
\begeq\label{3.12}
\lim_{L\to\infty}\frac{1}{2L}\int_{-L}^{L}\big|\gamma_L(x)\big|^2dx=
\frac{1}{4\pi^2}\sum_j \big|\Delta\sigma(q_j)\big|^2,
\endeq
{\it {where  the sum runs over all the discontinuity points $q_j$s of $\sigma(q)$ and $\Delta\sigma(q_j)\equiv 
[\sigma({q_j}^+)-\sigma({q_j}^-)]$.   }}\\   
{\rm {If $\eta(x)\in\cWp$, function $\sigma(q)$ is 
absolutely continuous by {\bSb}\  and all the $\Delta\sigma(q_j)$s 
vanish. Thus, the previous theorem becomes (II, pag. 181)}}:\\   
{\bf{P.13 -}} 
 {\em {if   $\eta(x)\in\cWp$, one has}}
\begeq\label{3.12a}
\lim_{L\to\infty}\frac{1}{2L}\int_{-L}^{L}\big|\gamma_L(x)\big|^2dx=0. 
\endeq
This theorem implies an interesting physical property  that was not explicitly noted by Wiener, namely:  for any amorphous system, $|\gamma(x)| $ is  
$O(x^a)$ with $a< 0$ as $x\to\pm\infty$. 
This remark proves the claim, reported in SAS textbooks and justified on the 
basis of the only  model worked out by  Debye, Anderson and Brumberger 
(Debye \etal, 1957), that the CF of any amorphous system vanishes at 
very large distances. \\
A further result concerns a stronger version of the inversion of relation (\ref{3.7}). 
In fact, theorem 34 of Wiener (II, pag. 182) states that:\\   
{\bf{P.14 - }} {\em {if $\eta(x)\in\cW$, then $\gamma(x)$ can be 
expressed as a Stieltjes-Lebesgue Fourier transform in so far the 
following equality}}
\begeq\label{3.13}
\gamma(x)=\frac{1}{2\pi}\int_{-\infty}^{\infty}e^{-ixq}d\sigma(q)
\endeq
{\em{ is almost everywhere true}}.\\
If $\eta(x)\in\cWp$, $\sigma(t)$ is absolutely continuous. Then  one 
can write $d\sigma(t)=\sigma'(t)dt$ and the above relation converts into  
\begeq\label{3.13a}
\gamma(x)=\frac{1}{2\pi}\int_{-\infty}^{\infty}e^{-iqx}\sigma'(q)dq=
\frac{1}{2\pi}\int_{-\infty}^{\infty}e^{-iqx}I(q)dq
\endeq 
owing to relation  (\ref{3.9b}). 
One can then state:\\     
{\bf{P.15 - }} {\em {if $\eta(x)\in\cW_p$, according to (\ref{3.13a}),  $\gamma(x)$ is the FT of the scattering intensity,}}\\ 
a property  not reported by Wiener because it follows from assumption 
{\bC}, \ie\ from the a.e. continuity of $I(q)$.  
Besides, relation (\ref{3.13a}) coincides with the $\epsilon\to 0$ limit, 
taken inside in the integral, of equation (\ref{3.7b}). Hence, if $\eta(x)\in\cWp$, the $\epsilon\to 0$ limit can be exchanged with 
the integral on the rhs of (\ref{3.7b}). Moreover {\bf P.15} and 
{\bf P.11} allows us to get a stronger bound on the asymptotic behavior 
of the CF at large $x$s. In fact, {\bf  P.11} implies that 
$I(q)=O(q^\alpha)$ with $\alpha<-1$ as $q\to\pm\infty$ so that 
$I(q)\in L_2$.  Then, one can apply the Plancherel theorem to (\ref{3.13a})  
and one obtains 
\begeq\label{3.13b}
\int_{-\infty}^{\infty} I^2(q)dq=2\pi
\int_{-\infty}^{\infty}\big|\gamma(x)\big|^2dx.
\endeq 
The last integral clearly implies that $|\gamma(x)|=O(x^{\beta})$ with 
$\beta<-1/2$ at large $x$s. This bound is stronger than the one 
obtained below  {\bf P.13} and implies that $\gamma(x)\in L_2$.  
Hence, the quite important property (not reported by Wiener): \\  
{\bf{P.16 - }} {\em {if $\eta(x)\in\cW_p$, the associated CF $\gamma(x)$ and scattering intensity $I(q)$ are $L_2(-\infty,\infty)$ summable functions.}}\\
This property deserves  some words of comment. SAS textbooks 
implicitly assume that the CFs of amorphous samples are continuous 
$L_2$  functions and consequently consider observed scattering intensities 
as their FTs. Besides, they also implicitly assume that 
observed scattering intensities obey  the Porod  invariant relation. This 
last condition implies that, at large $q$s,  $I(q)=O(q^{a})$ with $a<-1$. 
From the last relation follows that $I(q)^2=O(q^{-2a})$ at large $q$ so  
that the  $I(q)^2$  decrease at large $q$s is faster than the one required 
for the function to belong to $L_2$. Then, a theorem by Plancherel (1915) 
ensures us that the FT that relates $I(q)$ to $\gamma(x)$ 
converges almost everywhere, as specified in equation (\ref{3.13a}),  
and not only in the mean. One concludes that the two assumptions 
understood in SAS textbooks imply the validity of the conditions required 
by statement {\bSb} while  our analysis has rigorously shown  that 
the two assumptions follow from  Wiener's results and assumption {\bC}.\\  
The properties of the functions belonging to $\cWp$ are further clarified by theorem 33 of Wiener (II, pag. 182). This states that:\\     
{\bf{P.17 - }}{\em { given a CF $\gamma(x)$ generated by a SDF $\eta(x)\in \cW$, put}}
\begeq\label{3.14}
\gamma_1(x)=\gamma(x)-\frac{1}{2\pi}\sum_j \Delta\sigma(q_j) e^{iq_jx},
\endeq
{\em {where the $\Delta\sigma(q_j)$s and the $q_js$ have been defined 
below equation (\ref{3.12}), then $\gamma_1(x)$ obeys  sum-rule (\ref{3.12a})}}, \ie\\
\begeq\label{3.15}
\lim_{L\to\infty}\frac{1}{2L}\int_{-L}^{L}\big|\gamma_1(x)\big|^2dx=0. 
\endeq 
It was already remarked that the last condition implies that 
$\gamma_1(x)\to 0$ as $x\to \pm\infty$. It is also noted that, 
owing to equation (\ref{3.11b}), the series present on the rhs 
of (\ref{3.14}) is absolutely convergent so that the function
\begeq\label{3.16} 
\cA(x)\equiv \frac{1}{2\pi}\sum_j \Delta\sigma(q_j) e^{iq_jx}
\endeq 
is a function continuous, bounded and defined throughout 
$(-\infty,\infty)$. Besides it is a {\em {uniform almost periodic}} function 
(UAPF) (for the definition see: II section 24) in so far 
it obeys the conditions required by Bohr's theorem (II, pag. 186), 
namely:\\      
{\bf{P.18 - }}{\em {a function $\cP(x)$ is a   UAPF iff }}
\begeq\label{3.17}
{\widehat{\cP}}(\alpha)\equiv
\lim_{L\to\infty}\Bigl\{\frac{1}{2L}\int_{-L}^{L}\cP(x)e^{i\alpha x}dx\Bigr\}
\endeq
{\em{exists for any real $\alpha$, it is equal to zero except  
at a discrete set of $\alpha$ values $(\alpha_1,\, \alpha_2,\ldots)$ 
where it takes the finite non null values $A_j=\widehat{\cP}(\alpha_j)$  
(with $j=1,2,\ldots$) and it finally results}}
\begeq\label{3.18}
\lim_{N\to\infty}\lim_{L\to\infty}\Bigl\{\frac{1}{2L}
\int_{-L}^{L}\big|\cP(x)-\sum_{j=1}^N A_j e^{-i\alpha_j x}\big|^2dx\Bigr\}=0, 
\endeq 
that is also equivalent to 
\begeq\label{3.18a}
\lim_{L\to\infty}\Bigl\{\frac{1}{2L}
\int_{-L}^{L}\big|\cP(x)|^2dx\Bigr\}=\lim_{N\to\infty}\sum_{j=1}^{N} \big|A_j\big|^2. 
\endeq 
Even though equations (\ref{3.17}) and (\ref{3.18a}) look respectively 
similar to the FT definition and the Parseval equality, a quite important difference must be noticed: the presence  of the diverging denominator 
$2L$,  present  neither in the standard FT definition nor in the standard Parseval relation. [For this reason we used the caret instead of the tilde on 
the lhs of (\ref{3.17}).] This difference arises from the fact that the 
standard FT definition and Parseval equality deal with functions that are quadratically summable while the $\cP(x)$, present in equations 
(\ref{3.17}) and (\ref{3.18a}), does not belong to $L_2(-\infty,\infty)$, 
it only being bounded. To be more explicit, a UAPF $\cP(x)$ is 
characterized by the property that the integral on the rhs of 
(\ref{3.17}) is $O(L)$ only at the discrete set of  the $\alpha_j$ 
values while, at remaining $\alpha$s,  it is $O(L^a)$ with $a<1$.    
By { {\bf P.17}} and the fact that function $\cA(x)$ defined by (\ref{3.16}) 
is a UAPF,  it follows that  {\bf{P.17}} can be restated by saying that\\ 
{\bf{P.19 - }} {\em if a CF does not go to zero at large distances, it deviates from the null value by the UAPF defined by (\ref{3.16})}.\\ 
It is useful to report a further property of UAPFs (see lemma $37_{15}$ 
at pag. 196 of II), namely:\\
{\bf{P.20 - }} {\em{if $\cP(x)$ is a UAPF, then there exists a 
discrete set of positive numbers $A_j$ and real numbers $\alpha_j$ 
such that  $\sum_j A_J<\infty$ and }}
\begeq\label{3.19}
\gamma_{\cP}(x)=\lim_{L\to\infty}\frac{1}{2L}
\int_{-L}^{L}\cP(x+y){\overline{\cP(y)}}dy=\sum_j A_j e^{i\alpha_j x}.
\endeq
This property states that a UAPF generates a CF that also is a 
UAPF and that, by the very definition of uniform almost periodicity,  cannot vanish at very large distances. This allows us to better characterize 
the functions that belong to $\cWp$ by stating that:\\ 
{\bf{P.21 - }} {\em{the functions that belong to $\cWp$ can neither be 
UAPFs nor contain a uniform almost periodic contribution}} (UAPC).\\
In fact,  if $\eta(x)\in\cWp$ the resulting CF vanishes at large distance,  
while if $\eta(x)$  is uniform almost periodic,  owing to property {\bf P.20}, 
the generated  CF does not asymptotically vanish. To prove the second 
part of the statement, we first show how $\cP_{\eta}(x)$, the UAPC  
present in a $\eta(x)\in\cW$, can be singled out.  One first evaluates the associated function, depending on $\alpha$, according to (\ref{3.17}), \ie
\begeq\label{3.20}
{\widehat\eta}(\alpha)=
\lim_{L\to\infty}\Bigl\{\frac{1}{2L}\int_{-L}^{L}\eta(x)e^{i\alpha x}dx\Bigr\}
\endeq 
and one looks for the $\alpha$ values, again denoted by $\alpha_j$, such 
that $A_j\equiv {\widehat\eta}(\alpha_j)\ne 0$. If these values exist, then 
one puts 
\begeq\label{3.21}
\cP_{\eta}(x)\equiv\sum_j A_je^{i\alpha_j x}\quad {\rm and}\quad  
\eta_1(x)\equiv\eta(x)-\cP_{\eta}(x). 
\endeq 
By construction, function $\eta_1(x)$ contains no UAPC and 
$\cP_{\eta}(x)$ is the UAPC present in $\eta(x)$. It is straigthforwad now 
to show that the CF generated by $\eta(x)$ is equal to the sum of the 
CFs respectively generated by $\eta_1(x)$ and $\cP_{\eta}(x)$ because  
the two "interference" integrals 
\begeq
\frac{1}{2L}\int_{-L}^{L}\eta_1(x+y){\overline{\cP_{\eta}(y)}}dy\quad{\rm and}\quad 
\frac{1}{2L}\int_{-L}^{L}\cP_{\eta}(x+y){\overline{\eta_1(y)}}dy\nonumber
\endeq
vanish in the $L\to\infty$ limit. Since the CF of $\cP_{\eta}(x)$ is 
a UAPF that does not asymptotically vanish, one concludes that, if $\eta(x)\in\cWp$,  $\eta(x)$ cannot contain a UAPC  
and the proof of {\bf P.21} is completed.\\
We remark that, if a function $\eta(x)$ belongs to $\cWp$, its 
mean value $\langle \eta\rangle$ must necessarily vanish  because  
$\widehat\eta(\alpha)=0$ for any $\alpha$ and then  $\langle \eta\rangle=\widehat\eta(0)=0$.\\
So far function $s(q)$, defined by (\ref{3.3a}), was apparently put aside. 
It's time now to discuss some of its properties.  The first concerns 
its asymptotic behavior at large $q$'s that is partly specified by theorem  
28 by Wiener (II, pag. 160) which states that:\\
{\bf{P.22 - }} {\em {if $\eta(x)\in\cW$, it will belong to $\cW'$ iff the 
relation}}\\
\begeq\label{3.13c}
\lim_{L\to\infty}{\overline{\lim_{\epsilon\to 0}}}
\frac{1}{4\pi\epsilon}\Biggl[\int_L^{\infty}+
\int_{-\infty}^{-L}\Biggr]|s(q+\epsilon)-s(q-\epsilon)]|^2dq=0
\endeq
{\em holds true.}\\
Since $\cWp$ is a subset of $\cW'$,  it follows that 
the  $s(q)$ associated to any $\eta(x)\in\cWp$  obeys to (\ref{3.13c}).\\   
Another property is the following:\\ 
{\bf{P.23 - }} {\em{if $\eta(x)$ contains the periodic contribution 
$e^{i\alpha x}$, then $s(q)$ jumps at $q=-\alpha$.}}\\      
In fact, putting $\eta(x)=e^{i\alpha x}$ within (\ref{3.3}) and (\ref{3.4}) 
one finds  
\begin{eqnarray}\label{3.22}
&&s_L(q)\equiv s_{1,L}(q)+s_2(q)=
-2\bigl\{ {\rm si}(\alpha) -{\rm si}\bigl[L (\alpha + q)\bigr] \bigr\} 
\end{eqnarray}
where ${\rm si}(\cdot)$  denotes the sine integral 
function  (Gradshteyn \& Ryzhik, 1980, sect. 8.2). The $L\to\infty$ 
limit of $s_L(q)$ yields 
\begeq\label{3.23}
s(q)=\begin{cases}
-2{\rm si}(\alpha)-\pi&\text{if}\quad q<-\alpha,\\
-2{\rm si}(\alpha)&\text{if}\quad  q=-\alpha,\\ 
-2{\rm si}(\alpha)+\pi &\text{if}\quad  q>-\alpha,
\end{cases}
\endeq 
and the property is proved. \\
The integral relation (\ref{3.10}) implies that:\\  
{\bf{P.24 - }} {\em{the points of jumps of $s(q)$ also are points of 
jumps of $\sigma(q)$ and vice versa.}}\\
In fact, let  $q_0$ denote a point of jump of $s(q)$. On the left and 
right neighbouroods of this point, $s(q)$ will respectively behave as $a_l+b_l(q_0-q)^{\alpha_l}$ and $a_r+b_r(q-q_0)^{\alpha_r}$ with $\alpha_l>0$, $\alpha_r>0$  and $a_l,\, b_l,\,a_r,\,b_r$ 
complex numbers. Evaluate equation (\ref{3.10}) at 
$q=q_0-\tau$ and at $q=q_0+\tau$ with $\tau>0$. Subtracting the two 
results one finds
\begin{eqnarray}
&&4\pi\big\{[\sigma(q_0+\tau)-\sigma(q_0-\tau)]+[\sigma(-q_0+\tau)-\sigma(-q_0-\tau)]\bigr\}=
\nonumber \\
&&\limM_{\epsilon\to 0}\Bigl[\int_{q_0-\tau}^{q_0+\tau}
\frac{|s(t+\epsilon)-s(t-\epsilon)|^2}{\epsilon}dt+
\int_{-q_0-\tau}^{-q_0+\tau}\frac{|s(t+\epsilon)-s(t-\epsilon)|^2}{\epsilon}dt\Bigr].\nonumber
\end{eqnarray}
The above expression must be considered in the limit $\tau\to 0$. 
Having assumed $s(q)$ continuous at $q=-q_0$, the second integral 
vanishes in the limit $\epsilon\to 0$ and one is left with 
\begeq\label{3.24}
\lim_{\tau\to 0}[\sigma(q_0+\tau)-\sigma(q_0-\tau)]=\lim_{\tau\to 0}
\lim_{\epsilon\to 0}\int_{q_0-\tau}^{q_0+\tau}\frac{|s(t+\epsilon)-s(t-\epsilon)|^2}{4\pi\epsilon}dt.
\endeq 
The rhs can explictly be evaluated using the reported leading 
expressions of $s(q)$ and one finds 
\begeq\label{3.25}
\Delta\sigma(q_0)\equiv\sigma({q_0}^+)-\sigma({q_0}^-)=\frac{(a_r-a_l)^2}{2\pi}\equiv
\frac{\Delta s(q_0)^2}{2\pi}.
\endeq 
Hence, the jump of $s(q)$ at $q_0$ reflects into a jump of $\sigma(q)$ at 
the same $q$ value. On the contrary, if $\sigma(q)$ has a jump at $q_0$, 
the lhs of (\ref{3.24}) is finite. Consequently,  $s(q)$ cannot be continuous at $q_0$ otherwise the rhs of (\ref{3.24}) would be equal to 
zero  with a contradictory result.\\
It was already shown that SAS from amorphous samples requires that  SDFs $\eta(x)$ belong to $\cWp$. Thus, properties {\bf P.20} ensures that 
$\sigma(q)$ has no jumps and {\bf P.24} that the same happens for $s(q)$. 
This last property implies that\\   
{\bf{P.25 - }} {\em{If $\eta(x)\in\cW_p$, then the total variation of 
$s(q)$ over $(-\infty,\,\infty)$ is unbounded.}}\\     
(We refer to  sect. IV.2 of Kolmogorov \& Fomin (1980) for the definition 
of the total variation.)     
This property implies that $s(q)$ is divergent at some $q$ 
point or has infinitely many  oscillations (that are generally not  
periodic and, in amplitude, do not decrease too fast), or both.  The 
proof of {\bf P.25}  follows immediately from  another result by 
Wiener (II, pag. 146) that states:\\ 
{\bf{P.26 - }} {\em{Let $\eta(x)\in \cW$ and let $s(q)$ be defined as in equation (\ref{3.3a}). 
If $s(q)$ has bounded total variation over $(-\infty,\,\infty)$, then $\gamma(0)$  is equal to the sum of the jumps of $s(q)$ 
at its points of discontinuity. Consequently, if $s(q)$ is continuous, 
one necessarily has that $\gamma(0)=0$.}} \\  
In fact, since any $\eta(x)\in\cW_p$ generates a non null CF 
as well as a continuous $s(q)$, the boundedness condition of the 
total variation of $s(q)$  must necessarily be violated otherwise, 
according to {\bf P.26}, the CF would be null contradicting the assumption.\\ 
This section can be summarized stating that the results of Wiener,  
combined with assumption \bC, require that  a function $\eta(x)$  
must belong to $\cWp$ for it to may be considered the SDF of 
an amorphous sample. In fact,  the condition that $\eta(x)\in \cWp$
ensures, by {\bSb},  that  the relevant CF exists and is continuous throughout $(-\infty,\,\infty)$ and that the associated integrated scattering intensity $\sigma(q)$, defined by equation (\ref{3.6}), exists, is non-decreasing and absolutely  continuous  throughout $(-\infty,\,\infty)$. Besides, the knowledge of $\sigma(q)$ uniquely determines, {\em via} equation (\ref{3.7b}) or (\ref{3.13}), $\gamma(x)$ that results to be a continuous and  an $L_2$ function so that $I(q)$ can simply be 
expressed as its FT.  At the same time, the function 
$s(q)$, defined by equations (\ref{3.3c}) and (\ref{3.3a})-(\ref{3.4}), 
also  exists in the mean, is related to $\sigma(q)$  by (\ref{3.10}), 
has no jumps (see {\bf P.24}) and and unbounded total variation 
(see {\bf P.25}).  \\ 
\subsection*{4. Another way of characterizing physical SDF}
We have just said that the property crucial for the existence of $I(q)$ is 
that $\eta(x)$ belongs to $\cWp$, \ie\ condition {\bSb}. We show now 
that,  if this condition is obeyed,  statement \bSa\ also holds true. In fact, 
{\bf S$_2$} implies that $I(q)$ exists, is non-negative  and a.e. continuous.  
Equation (\ref{1.8}), adapted to the one dimensional case, implies that 
\begeq\label{4.1}
I(q)=\lim_{L\to\infty}I_L(q)=\lim_{L\to\infty}
\frac{\big|\teta_L(q)\big|^2}{2L},
\endeq
and from this relation the existence of $I(q)$ implies that of  the limit 
\begeq\label{4.2}
\lim_{L\to\infty}\Big|\frac{\teta_L(q)}{\sqrt{2L}}\Big|.
\endeq
 At the same time, the other condition, involved in statement \bSa\ 
[\ie\ equation (\ref{1.9})], in the one dimensional case takes the form
\begeq\label{4.2e}
\int_{\infty}^{\infty}\Biggl[\lim_{L\to\infty}\frac {\Big|\teta_L(q)\Big|^2}{{2L}}\Biggr]dq=
\int_{-\infty}^{\infty} I(q)dq=(2\pi)\langle |\eta|^2\rangle=2\pi\gamma(0)  
\endeq 
that is ensured by  {\bf P.10}, and the proof is completed. On the contrary, 
assume  that statement \bSa\ be true.   It follows 
that $\eta(x)\in\cWp$. In fact, one has 
\begin{eqnarray}\label{4.3}
&&\lim_{L\to\infty}\frac{\big|\tn_L(q)\big|^2}{2L}=
\int_{-\infty}^{\infty}e^{iqt}dt \lim_{L\to\infty}\frac{1}{2L}
\int_{-\infty}^{\infty}\eta_L(y+t){\overline{\eta_L(y)}}dy=\nonumber\\
&& \quad\quad \quad 
\int_{-\infty}^{\infty}\bigl[\lim_{L\to\infty}\gamma_L(t)\bigr]e^{iqt}dt=
\int_{-\infty}^{\infty}\gamma(t)e^{iqt}dt 
\end{eqnarray}
(where the limit exchange is ensured by the finite supports of the 
integrand functions). Then, the assumed existence of the limit on the 
lhs  ensures the existence of the FT and of $\gamma(x)$. Besides, 
owing to  equation (\ref{4.2e}), one has that $I(q)=O(q^\alpha)$ 
with $\alpha<1$ at large $q$s so that $I(q)\in L_2$ and, by the Plancherel 
theorem, one finds that $\gamma(x)\in L_2$.  Recalling {\bf S$_2$} and 
{\bf P.15}  one concludes  that  $\eta(x)\in \cWp$.\\   
In conclusion, we have  two equivalent ways for characterizing 
physical SDFs: either one requires that $\eta(x)\in\cWp$ or one 
requires that $\teta_L(q)$ obeys \bSa.  The first way is mainly based 
on the CF detrmination and its subsequent FT  while the second on the $L\to\infty$ limit of the FT of the SDF. \\ 
For completeness, we also recall a claim by Schuster (1906) [see sect. 2 
of I], according to which "the modulus of the FT of a random function, evaluated over a domain of size $L$, fluctuates around a mean value which increases as $L$ or as $\sqrt{L}$ depending on whether the function 
does or does not contain a periodic contribution".  It is clear that 
the assumed existence of limit (\ref{4.2}) coincides with Schuster's 
statement because  the SDF of any 
amorphous sample contains no UPAC responsible for an $O(L)$ 
contribution. However, it must also be said that Schuster claim 
is not correct because it rules out the possible existence of functions 
whose FTs's moduli increase as $L^{\alpha}$ with $\alpha\ne 1/2$ or $1$ 
as the integration domain size $L$ increases.  Actually, these functions 
do exist. An example due to Mahler (1927) is  reported at pag.s 
203-209 of I.  \\ 
Unfortunately no explicit model of  a SDF $\eta(x)$ obeying  \bSa\ 
is as yet known, and one might wonder whether $\cWp$ be not a void set.  
On physical ground there is no room for this doubt.   In the following subsections we numerically analyze two $\eta(x)$ models and the 
results indicate that they could be considered physical SDFs. 
\subsubsection*{4.1 The Wiener model}
We begin by recalling  that Wiener in II (pag.s 151-153) considered a model 
(hereinafter referred to as Wiener's) that leads to the continuous and $L_2$ 
CF
\begeq\label{4.4a}
\gamma(x)=
\begin{cases}
1-|x|&\text{if}\quad  |x|\le 1,\\
0 &\text{if}\quad  |x|\ge 1.
\end{cases}\endeq
The model is defined as 
follows.  Consider an irrational number $\alpha$ such that $0<\alpha<1$ 
and let $0.a_1a_2a_3a_4\ldots$ denote its binary representation, \ie\ $\alpha=\sum_{k=1}^{\infty} a_k/2^k$. Clearly, each 
$a_n$ is either equal to 1 or to 0, and the binary representation is not 
periodic for the assumed irrationality of $\alpha$.  
The associated SDF $\eta_{\alpha}(x)$ is defined, in arbitrary units 
(a.u.), as 
\begeq\label{4.4b}
\eta_{\alpha}(x)\equiv
\begin{cases}
2a_{2n+1}-1,\quad\text{if}\quad n<x\le n+1, \ \ \quad n= 0,1,2,\ldots\\
2a_{2n}-1,\ \ \ \quad\text{if}\quad -n<x\le 1-n, \quad n=-1,-2,\ldots
\end{cases}\endeq
where the length units are $u$. 
More explicitly, the $a_m$s with $m$ even and $m$ odd respectively determine the values of $\eta_{\alpha}(x)$ in the unit intervals set on 
the left and on the right of the origin, and  the codomain of 
$\eta_{\alpha}(x)$ is formed by the two element set $\{-1,\,1\}$.  
These facts, combined with the unit length of the intervals, yield the relation 
\begeq\label{4.5}
\gamma(x)=(n+1-x)\gamma(n)+(x-n)\gamma(n+1)\quad {\rm with}
\quad 0\le n\le|x|\le (n+1),
\endeq 
which shows that $\gamma(x)$ is fully determined once its values at  
all the integer $x$s have been determined. (For notational simplicity 
the dependence of $\gamma(\cdot)$ on $\alpha$ is 
omitted.)  Since $\eta_{\alpha}^2(x)\equiv 1$, it follows that 
$\gamma(0)=1$ and one is left with the determination of the 
$\gamma(n)$s with $n\ne 0$.  Borel (1909) established the property 
that, considered the binary representation of an irrational 
$\alpha\in (0,1)$, the probability that each $a_n$ be equal to 1 
or 0 is $1/2$.  On this probabilistic ground Wiener\footnote{
Actually, in evaluating the $N\to\infty$ limit of $\gamma_N(n)\equiv
\frac{1}{2N}
\sum_{m=-N+1}^N \eta_{\alpha}(n+m)\eta_{\alpha}(m)$, 
Wiener assumed that the probability that each term of the sum be 
equal to 1 or -1 is equal to $1/2$, which looks an assumption 
stronger than Borel's, which only states that the probability that each 
$\eta_{\alpha}(m)$ be equal to 1 or -1 is $1/2$.
} 
proved that $\gamma(n)=0$ if $n\ne 0$ and, from this result, 
the reported (\ref{4.4a}) expression of $\gamma(x)$ immediately 
follows by (\ref{4.5}).  This conclusion, however, cannot be 
considered equivalent to say that each $\eta_{\alpha}(x)$, 
associated to an $\alpha\in(0,\,1)$ by the above expounded 
procedure, certainly yields (\ref{4.4a}) as CF. The last statement, 
on a physical ground, looks rather unlikely. To make this point 
clear  we observe that  Wiener models can be interpreted 
as particulate two phase models of the M\'ering-Tchoubar (1968) 
kind. The particles correspond to the intervals where $\eta(x)=1$ 
and the voids to those where $\eta(x)=-1$. It is noted that  the 
particles and the voids can have arbitrary lengths depending on 
the $\alpha$ value. For instance, if  $a_1=a_3=a_5=1$ and $a_7=0$, 
we have a particle having the left end at the origin and the right
 end at $x=3$ and its length is equal to 3. 
Any one-dimensional amorphous model of the M\'ering-Tchoubar 
is fully defined by assigning the lengths $d_k$s and $z_k$s of all 
the particles and  voids contained in it. Index $k$ is assigned by 
setting the origin at the left end of an arbitrarily chosen particle 
and assigning the value $k=1$ to this particle. Then $z_1$ is the 
length of the void immediately on the right of particle 1, $d_2$ is 
the length of the particle next to  the right border of the void 1. 
Iterating this procedure all the $d_k$s and $z_k$s with $k\ge 1$ 
are uniquely defined. To define the $d_k$s and $z_k$s with 
negative $k$ values, one sets $z_{-1}$ equal to the length of 
the void whose right end coincides with the left
end of particle 1. Then $d_{-1}$ is the length of the particle 
next to and on the left of the void -1 just defined and, iterating 
the procedure,   all the $d_k$s and $z_k$s with $k\le -1$ are 
uniquely defined. 
[The value $k=0$ is of course excluded.] \\
We characterize now the M\'ering-Tchoubar models that 
are of the Wiener type. Consider the two sets of values 
$\{d_k|\, k=\pm1,\pm2,\ldots\}$ and 
$\{z_k|\, k=\pm1,\pm2,\ldots\}$. Let ${d}$ denote the greatest 
lower bound of the $d_k$s and ${ z}$ that 
of the $z_k$s.  If $d=z$ and $d>0$ and if, whatever $k$,  
$d_{k}/d$ and $z_k/z$ are integer numbers then, setting the 
length unit equal to $d=z$, the considered M\'ering-Tchoubar 
model is of the Wiener type because it exists a value of $\alpha$ 
that reproduces the  considered M\'ering-Tchoubar model. In 
fact, to determine $\alpha$,  one proceeds as follows. Since 
$d_1/d=n_1$ (with $n_1$ integer), one sets $a_1=a_3=\ldots=a_{2n_1+1}=1$ and, then, since 
$z_1/z=m_1$ (with $m_1$ integer) one sets $a_{2n_1+3}=a_{2n_1+5}=\ldots=a_{2n_1+2m_1+1}=0$ and 
so on for the positive $k$s. For the negative $k$s, one starts 
from $z_{-1}/z=m$ (with $m$ integer) and one sets $a_2=a_4=\ldots=a_{2m}=0$. Then one considers 
$d_{-1}/d=n$ and one sets 
$a_{2m+2}=a_{2m+4}=\ldots=a_{2m+2n}=1$ 
and so on. In this way $\alpha$ is fully and uniquely 
determined. One concludes that the Wiener models associated 
to the $\alpha$s lying within $(0,1)$ coincide with the 
M\'ering-Tchoubar models consisting of particles and voids 
having lengths integer multiple of the same unit length. Since 
it looks physically unlikely that all these models have 
the same scattering behavior, we believe that equation 
(\ref{4.4a}) is only true for a subset of the irrational 
$\alpha$s belonging to $(0,\,1)$.   \\ 
It is convenient to report the FT expressions of 
$\eta_{\alpha,N}(x)$ and $\gamma_N(x)$ for the Wiener 
model (restricted to the interval $[-N,\,N]$). 
Given a particular $\alpha$, one first evaluates  the $a_k$s, 
for $k=1,\ldots,2N$, that determine its binary representation 
up to the $2N$th digit. Then, equation  (\ref{4.4b}) defines  
the associated SDF $\eta_{{\alpha},N}(x)$ in the interval  
$[-N,\,N]$ and the value of $\eta_{{\alpha},N}(x)$ in the 
$k$th unit interval will simply be denoted by $\eta_k$ 
with  $k=-N,(-N+1),\ldots,(N-1)$.  
The FT of $\eta_{{\alpha},N}(x)$ reads 
\begeq\label{4.7}
\teta_{{\alpha,N}}(q)=\frac{\sin(q/2)}{q/2}e^{iq/2}{\tilde F}_N(q),
\endeq
where we put 
\begeq\label{4.7b}
{\tilde F}_N(q)\equiv \sum_{k=-N}^{N-1}\eta_k e^{i qk}.
\endeq
The FT of $\gamma_N(x)$  is  simply related to the square modulus of (\ref{4.7}) as
\begeq\label{4.7a}
{\tilde\gamma}_{{N}}(q)=\frac{1}{2N}\frac{\sin^2(q/2)}{(q/2)^2}
\Big|{\tilde F}_N(q)\Big |^2,
\endeq 
and the scattering intensity takes the form 
\begeq\label{4.7AA}
I(q)=\lim_{N\to\infty}{\tilde\gamma}_{{N}}(q)=\frac{\sin^2(q/2)}{(q/2)^2}
\nu^2 (q)
\endeq 
where 
\begeq\label{4.7AB}
\nu(q)\equiv \lim_{N\to\infty} \Big|\frac{{\tilde F}_N(q)}{\sqrt{2N}}\Big |.
\endeq
According to Wiener, the CF does not depend on $\alpha$ 
and is given by (\ref{4.4a}).  The FT of this expression yields
\begeq\label{4.8}
I(q)={\tilde\gamma}(q)=2(1-\cos q)/q^2=\sin^2(q/2)/(q/2)^2. 
\endeq
From this expression follows that 
\begeq\label{4.8A}
\int_{-\infty}^{\infty}I(q)dq=2\pi
\endeq 
in agreement with equation (\ref{4.2e}) because for any $\eta_{\alpha}(x)$ 
it results that $\langle \eta_{\alpha}^2\rangle=1$. 
From (\ref{4.7a}), (\ref{4.7AB})  and (\ref{4.8}) follows that
\begeq\label{4.9}
\nu^2(q)=\lim_{N\to\infty}\frac{1}{2N}
\Big|{\tilde F}_N(q)\Big |^2=1,  
\endeq
where, having simplified the factor $\sin^2(q/2)/(q/2)^2$, the 
equality may not be true at $q=2n\pi$ with $n=\pm 1,\,\pm 2,\ldots$.
 In the limit $N\to\infty$,  expression (\ref{4.7b}) becomes a Fourier 
series and  it will not be 
restrictive to confine ourselves to the $q$ range 
$0\le q<2\pi$. Relation (\ref{4.9}) implies that, 
as $N\to\infty$,  
\begeq\label{4.10}
\Big|{\tilde F}_N(q)\Big| = \sqrt{2N}+o(\sqrt{2N}),
\quad 0\le q<2\pi,
\endeq
for almost all the   $\{\eta_k\}$ sets  associated to the binary 
representations of  the irrational $\alpha$s lying within 
$(0,\,1)$.  Setting $q=0$ in equation (\ref{4.7}) and using (\ref{4.10}), 
it follows that 
\begeq\label{4.11}
\big|\langle \eta_{\alpha}\rangle_L\big|=\frac{1}{2L}\Big|\int_{-L}^{L}\eta_{\alpha}(x)dx\Big|=
\frac{1}{2N}\Big|\sum_{k=-N}^{N-1}\eta_k \Big |\approx 
\frac{1}{\sqrt{2L}}.
\endeq 
\begin{figure}[h]
{\includegraphics[width=7.truecm]{{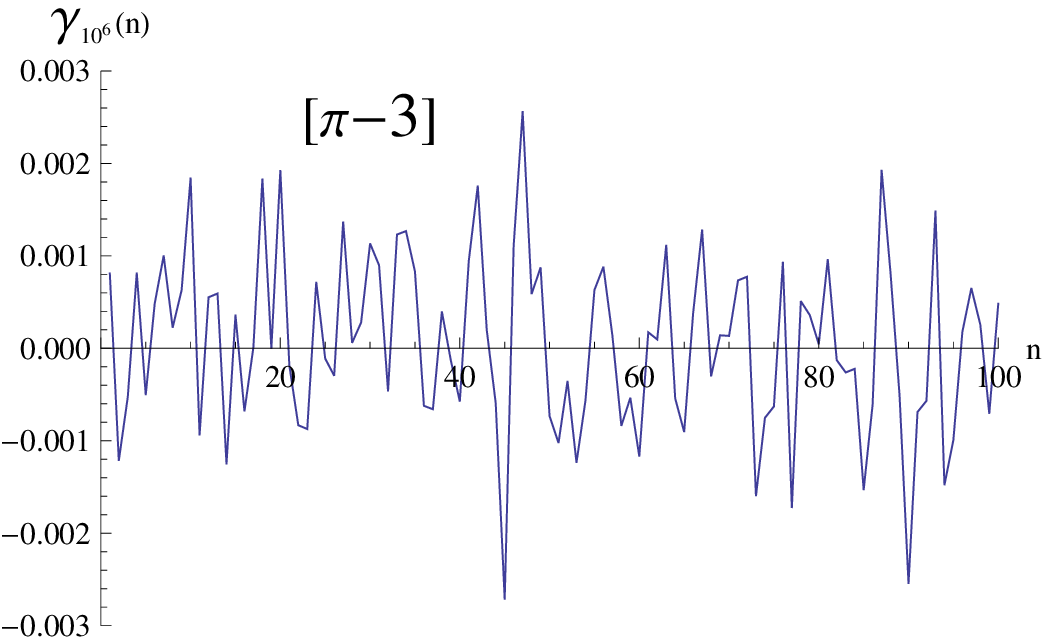}}}
{\includegraphics[width=7.truecm]{{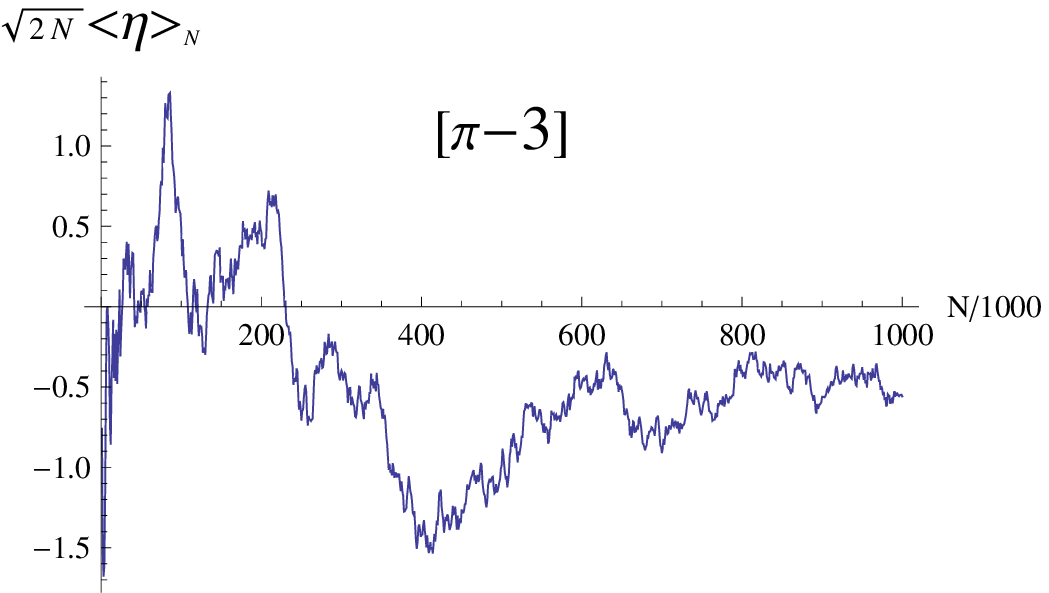}}}
{\includegraphics[width=7.truecm]{{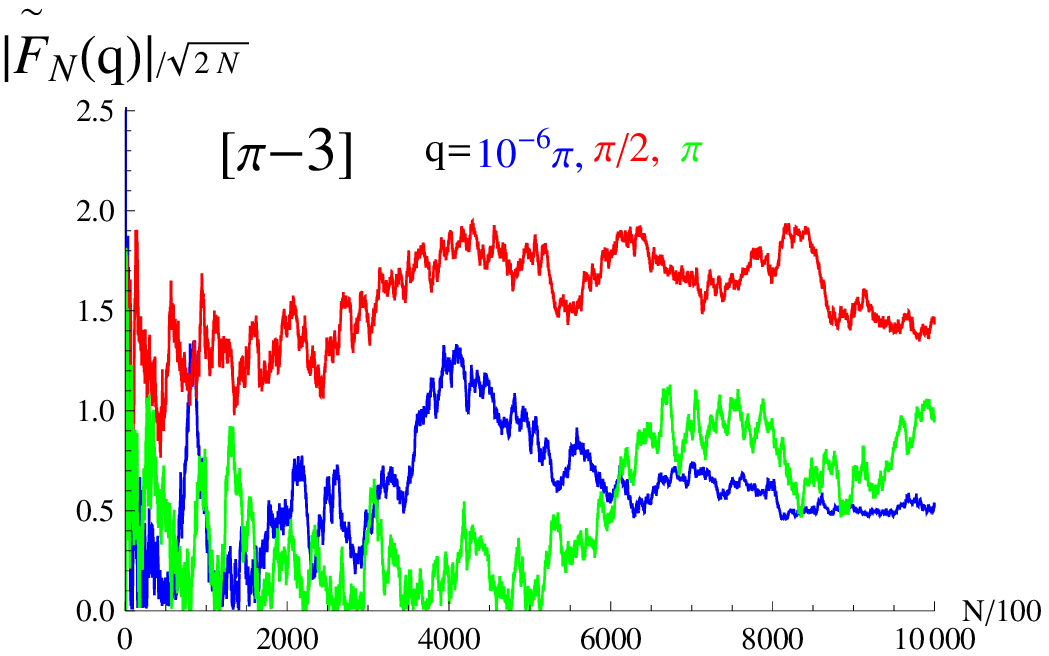}}}
{\includegraphics[width=7.truecm]{{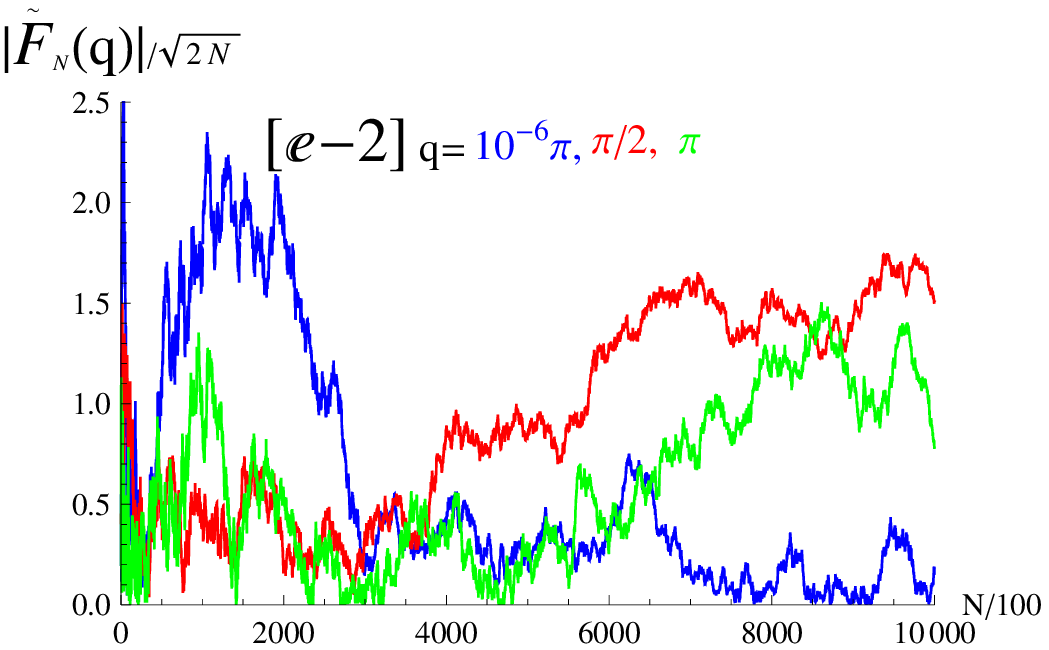}}}
\caption{\label{FigA1} {Plots of  some quantities relevant to the SDFs 
obtained by the Wiener models associated to  the transcendent  
numbers $(\pi-3)$  and $(e-2)$, numerically evaluated up to the
$2N=10^6$th binary digit.  [The horizontal axis label should read 
$2N/1000$ instead of the the reported $N/1000$.]}}
\end{figure}
This relation not only confirms that the mean value of the SDF $\eta(x)$, 
as expected, vanishes, but it also shows that,  as stated by Ciccariello 
(2017),   it vanishes as $|\nu|/\sqrt{2L}$ with $|\nu|=1$.  \\ 
We have tried to numerically test the validity of (\ref{4.11}) and (\ref{4.9}). 
To this aim we considered the cases $\alpha=(\pi-3)$ and $\alpha=(e-2)$ 
and  evaluated their  binary representations up to the $2N$th$=10^6$th 
binary digit.  
The corresponding $\eta_k$s are obtained by (\ref{4.4b}) which 
also yields the analytic expression of $\eta(x)$ for 
$-N<x\le N$. Knowing $\eta(x)$ it is numerically 
straightforward to evaluate the $\gamma_N(m)$s at some 
integer $m$ values, the mean value ${\langle \eta\rangle}_N$ 
of $\eta(x)$ over the interval $[-N,/,N]$ 
and $|{\tilde F}_N(q)|$, defined by (\ref{4.7b}), in terms of 
$N$ and $q$. Figure \ref{FigA1} shows the obtained results. 
The top left panel shows the $\gamma_N(m)$ values at 
the positive integers $m$ not exceeding 100 for the case 
$\alpha=(\pi-3)$. One sees that the values fluctuate around 
zero  and the largest deviations are  of the order of 
$1/\sqrt{2N}=10^{-3}$. The results for the case $\alpha=(e-2)$ 
are quite similar. Thus, they seem to confirm Wiener's conclusion 
that, in the $N\to\infty$ limit, all the $\gamma(m)$s are equal 
to zero whatever $\alpha$.  But, this conclusion might be 
hurried owing to the small range of the considered $m$s. 
[To enlarge this range, one should consider a much larger 
value of $N$ and this is not numerically easy owing to the 
exponential increase of the computation time.]  The top 
right panel plots $\sqrt{2N}{\langle \eta\rangle}_N$ 
{\em {versus}} $N$ in the range $1\le 2N\le 10^6$.  The tail 
of the curve shows an indication of a constant 
behavior and the same happens for the choice $(e-2)$, not 
shown in the figure. In the two cases, the constants look  to 
be $1/2$ and 0 and differ, therefore, from the value 1  
reported in equation  (\ref{4.11}). This fact indicates that 
either the  considered $2N$ value is not yet asymptotic or 
that Wiener equation (\ref{4.4a}) does not hold true for all 
the irrational $\alpha$s. The bottom left panel plots the 
quantity $|{\tilde F}_N(q)|/\sqrt{2N}$, evaluated for the 
$q$ values reported in the figure (the colors of the values 
are those of the associated curves), {\em versus} $N$. 
The blue curve practically coincides with the 
absolute value of the curve shown in the top right panel  
because the last curve is associated to $q=0$ which is close 
to $q=\pi/10^6$.  Finally, the bottom right panel shows the 
behavior of $|{\tilde F}_N(q)|/\sqrt{2N}$ for the same $q$ 
values, but it refers to the binary representation of $(e-2)$. 
It is not fully evident that  the tails of all the considered 
$|{\tilde F}_N(q)|/\sqrt{2N}$  curves  asymptotically show 
a constant behavior. If one inclines towards the 
affirmative answer, one would conclude that the constants 
are not equal to 1, as required by (\ref{4.10}), and the 
concern about Wiener result would be confirmed. Oppositely, 
one would only conclude that $2N$ is not yet sufficiently 
large to reach the asymptotic region where all the 
$|{\tilde F}_N(q)|/\sqrt{2N}$\,s coincide with 1. 
In any case, the important point is that the bottom 
panels of  Fig. \ref{FigA1} numerically show that the  
ratios $|{\tilde F}_N(q)|/\sqrt{2N}$ neither diverge nor tend 
to zero. They look  to tend to finite values,  depending on $q$, 
as required by (\ref{4.2}). According to \bSa, the resulting 
scattering intensity must obey (\ref{4.2e}). We already saw in 
(\ref{4.8A}) that  the sum rule  is  obeyed if 
if $\nu(q)=\lim_{N\to\infty}|{\tilde F}_N(q)|/\sqrt{2N}=1$. 
On the contrary, if  $\nu(q)$ is a periodic function of $q$, 
condition (\ref{4.2e}) converts into 
\begeq\nonumber 
\int_{-\infty}^{\infty}I(q)dq=\int_{-\infty}^{\infty}\frac{\sin^2(q/2)}{(q/2)^2}\nu^2(q)dq=2\pi. 
\endeq
We are certain that the last integral converges but the poor numerical 
accuracy in the $\nu(q)$ knowledge does not allow a reliable 
check of the relation.   
\subsubsection*{4.2 Another model}
\begin{figure}[h]
{\includegraphics[width=7.truecm]{{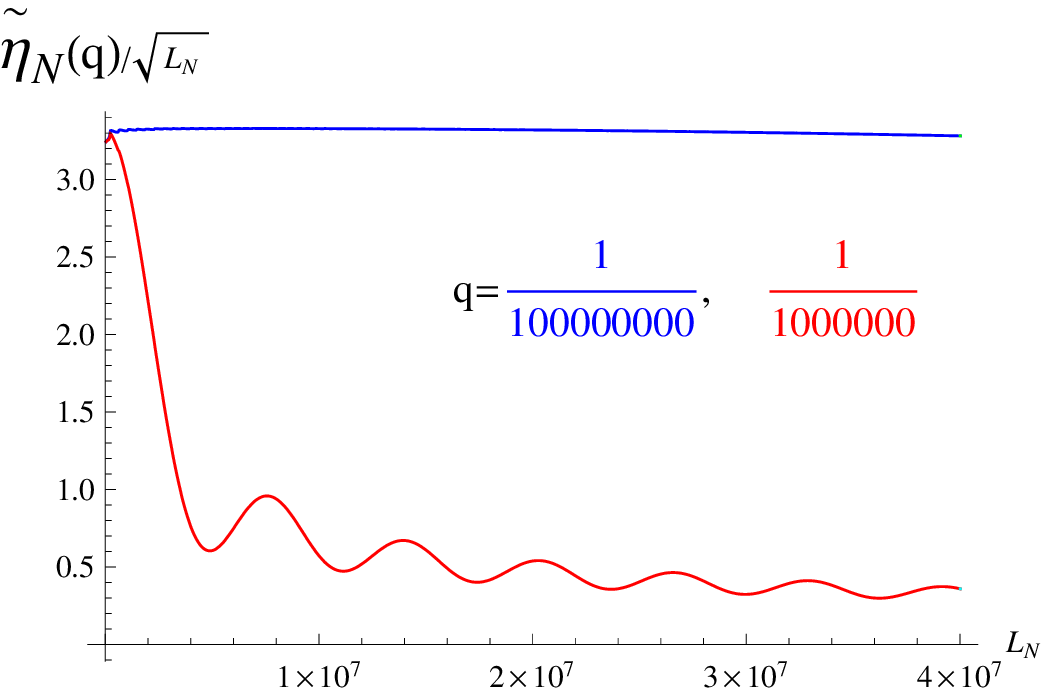}}}
{\includegraphics[width=7.truecm]{{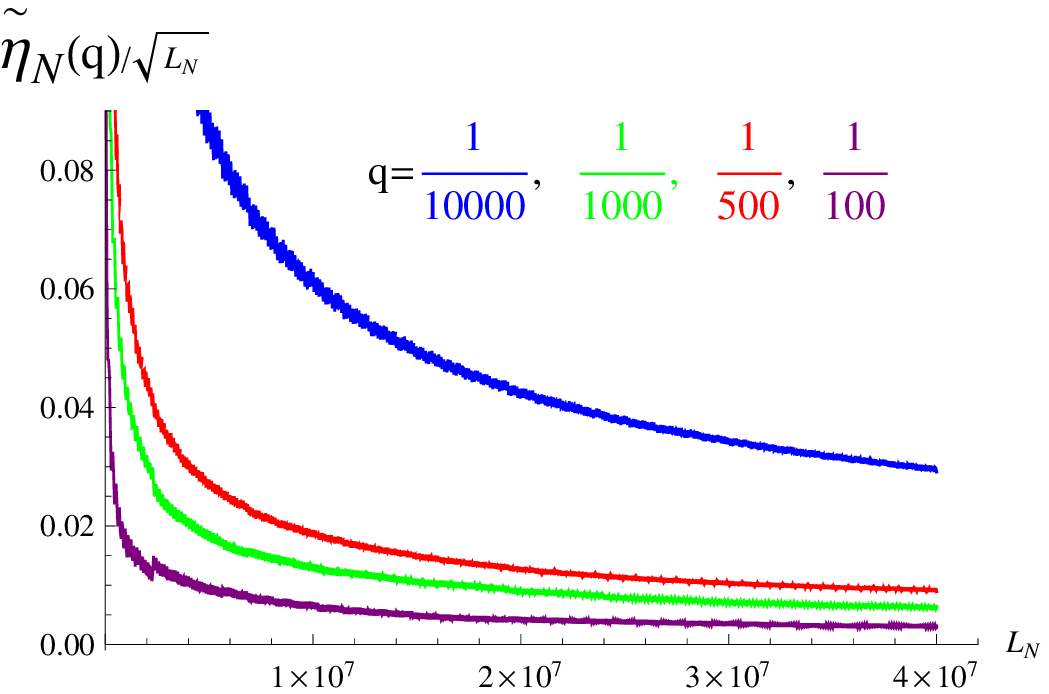}}}
{\includegraphics[width=7.truecm]{{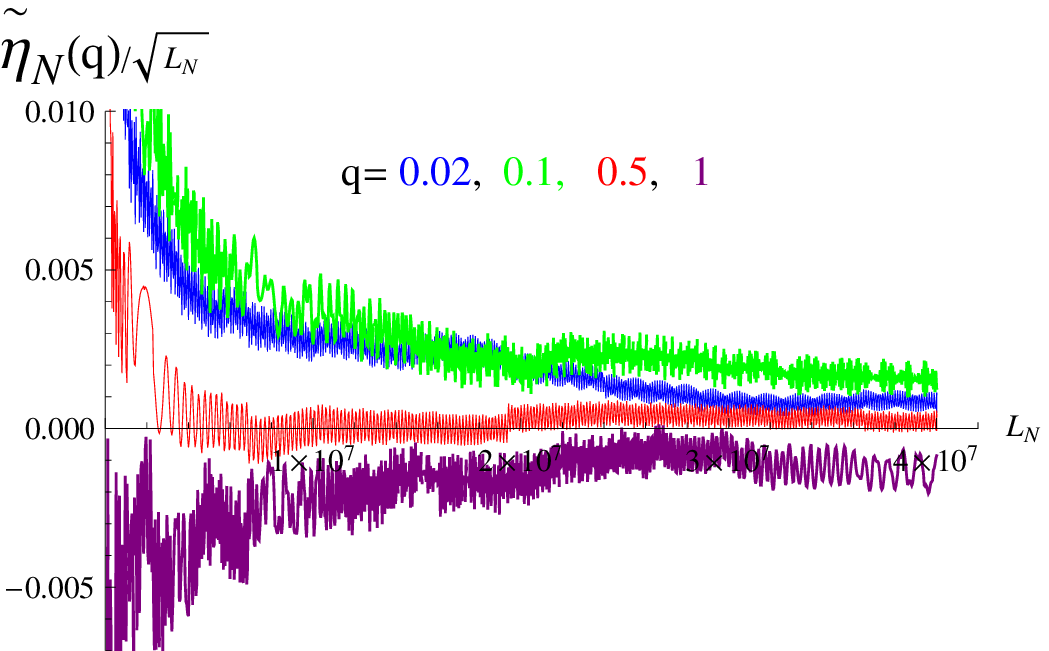}}}
{\includegraphics[width=7.truecm]{{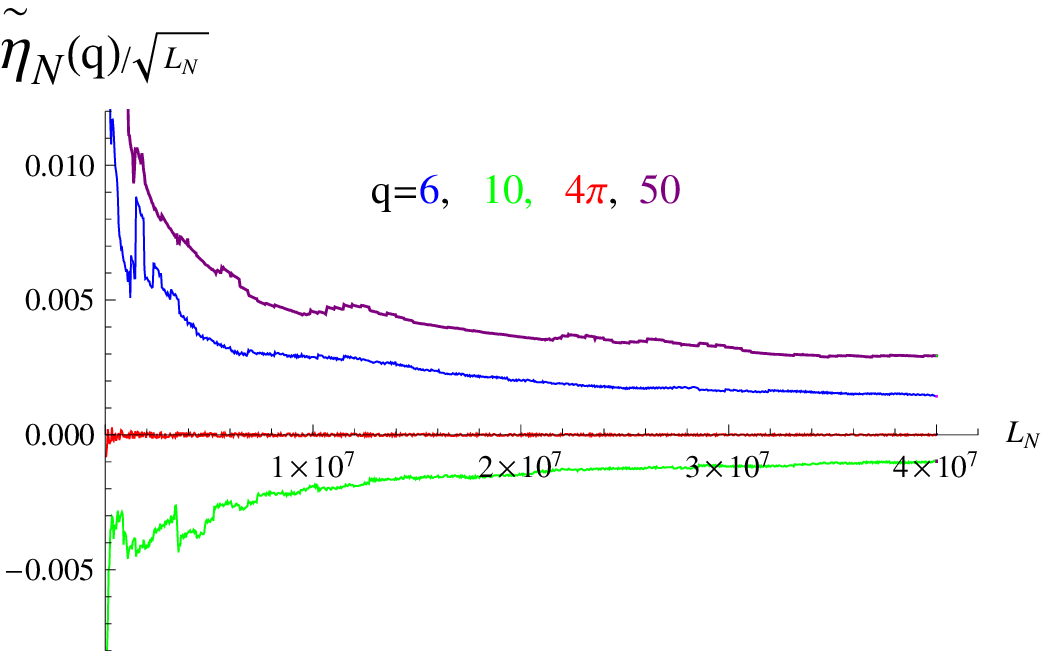}}}
\caption{\label{FigB3} {Plots of  $\teta_{N}(q)/\sqrt{L_N}$ (in a.u.) 
{\em versus} 
$N$ for different values of $q$ reported in each panel.  }}
\end{figure}

\begin{figure}[h]
\center{\includegraphics[width=7.truecm]{{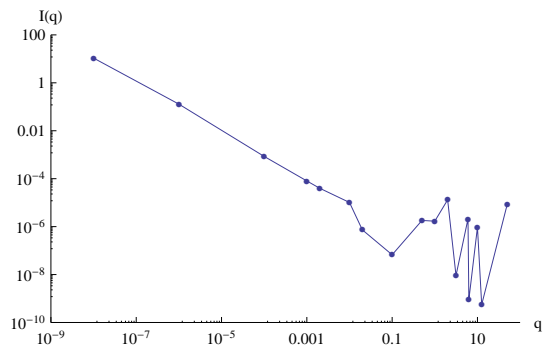}}}
\caption{\label{FigB4} {Log-log plot of the scattering intensity 
{\em versus} $q$ for the M\'ering-Tchoubar model defined by 
(\ref{4.20}) with $a=2,\, b=1,\, c=10^{-2}$. The intensity units 
are arbitrary and those of $q$ are $u^{-1}$, $u$ being the unit 
length used to define the particles' and voids' sizes. }}
\end{figure}

We numerically analyze now another model of the M\'ering-Tchoubar 
kind but not of Wiener's. We again assume  that the scattering 
densities of particles and voids are equal to 1 and -1, 
that particle 1 has the left border at the origin $O$ and that 
$\eta(x)$ is an even function. Thus, to fully characterize the 
model, it is sufficient to assign the lengths of particles and 
voids in the only region $x>0$. The lengths of the particles 
and the voids are assigned according to the following definitions
\begeq\label{4.20}
d_k\equiv a+\Delta_k,\quad  z_k\equiv a,\quad 
\Delta_k\equiv \frac{b}{\sqrt{c^2+k(1-\cos k)}},\quad k=1,2,\ldots
\endeq
where $a$, $b$ and $c$ are positive numbers, later specified. 
\begin{figure}[h]
\includegraphics[width=10.truecm]{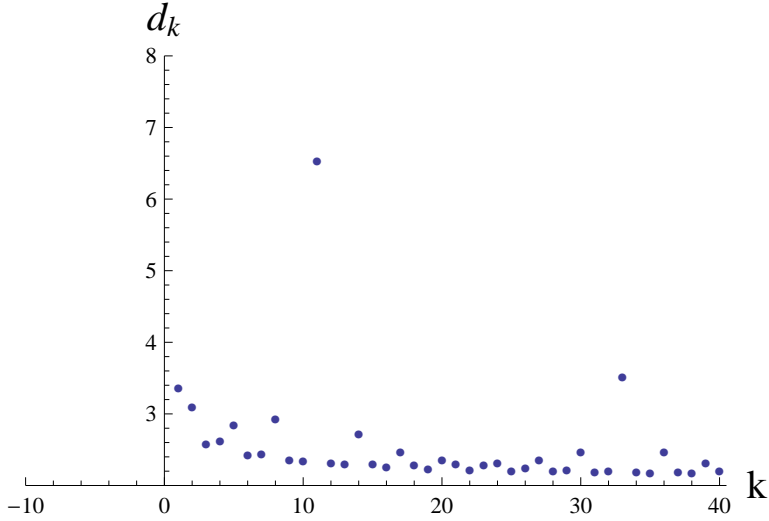}
\caption{\label{FigB1} {Plot of the first 40  $d_k$ values, in units $u$, for 
the case $a=2,\, b=1,\, c=1/100$. }}
\end{figure}
The greatest lower bounds of the $d_k$s and the $z_k$s are 
equal to $a$, but the ratios $d_k/a$ are not integers. This 
makes the model different from Wiener's. The essential 
ingredient is contribution $\Delta_k$ present in the $d_k$ 
definition.  Since $0\le (1-\cos x)\le 2$, 
it happens that $k(1-\cos k)$ is close to zero at  some particular $k$s 
that can also be arbitrarily large.  This property makes the 
$d_k$ values random, even though they obey the inequalities 
$a<d_k<a+b/c$. Besides, it ensures a stronger homogeneity 
in comparison to the case where contribution $\cos k$ is 
omitted in the $d_k$ definition,  because comparatively large 
$d_k$s can also be met at large $k$ values.  For illustration, 
the first 40  $d_k$ values for the case  
$a=2,\, b=1,\, c=10^{-2}$ are shown in figure \ref{FigB1}. \\ 
The particle $(\varphi_1)$ and void volume ($\varphi_2$) 
fractions are obtained evaluating the sums of particle and 
void lengths up to the $N$th void, \ie
\begeq\label{4.21}
D_N=\sum_{k=1}^N d_k=Na+\cD_N,\quad Z_N=Na,
\quad \cD_N\equiv\sum_{k=1}^N\Delta_k,
\endeq 
and taking the limit $N\to\infty$ of the appropriate ratios, \ie 
\begin{eqnarray}\label{4.22}
&&\varphi_1=\lim_{N\to\infty}\varphi_{1,N} \equiv \lim_{N\to\infty}\frac{Na+\cD_N}{2Na+\cD_N},\\
&&\varphi_2=\lim_{N\to\infty}\varphi_{2,M}\equiv
\lim_{N\to\infty}\frac{Z_N}{2Na+\cD_N}.\nonumber
\end{eqnarray}
Since $\varphi_{1,N}+\varphi_{2,N}=1$, the expected relation 
$\varphi_{1}+\varphi_{2}=1$ is obviously obeyed. A rough 
estimation of  the way $\cD_N$ depends on $N$ at large 
$N$s  can be obtained substituting, in the $\Delta_k$ 
definition, the factor $(1-\cos k)$ with a positive constant 
$ C$, representing a sort of mean value, and converting 
the $\cD_N$ expression into an integral. In this way one finds 
\begin{eqnarray}\label{4.24}
&&\cD_N\approx \sum_{k=1}^N\frac{b}{\sqrt{c^2+C k}}
\approx\int_0^N\frac{b}{\sqrt{c^2+C x}}dx=\nonumber\\
&&\quad\quad \frac{2b}{C}\bigl[\sqrt{c^2+C N}-c\bigr]\approx 2b(N/C)^{1/2}+o.
\end{eqnarray} 
This relation shows that $\cD_N$ linearly increases with 
$N^{1/2}$ and from (\ref{4.22}) one obtains that 
$\varphi_1=\varphi_2=1/2$. 
By equations (\ref{4.21}) and (\ref{4.24}) it is also possible 
to determine how the mean value of the considered SDF 
over the interval of length 
$L_N[\equiv(2Na+\cD_N)]$ behaves  as $N$ increases. 
\begin{figure}[h]
\includegraphics[width=10.truecm]{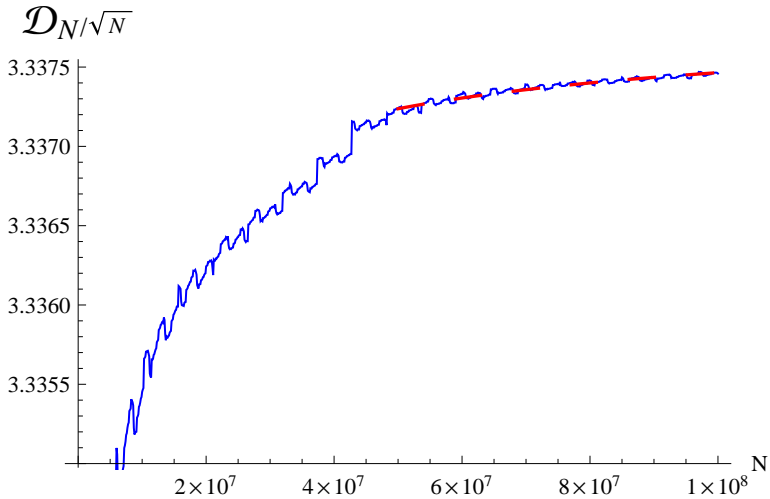}
\caption{\label{FigB2} {Plot of quantity $\cD_N$ (defined by (\ref{4.21}) and 
in units $u$)  
relevant to $N$ subsequent particle-void pairs and divided by $N^{1/2}$,  
versus $N$. The broken red curve approaches the constant value A=13.84 
(see text). }}
\end{figure}
One finds
\begeq\label{4.25} 
\langle \eta \rangle_{L_{N}}=\frac{1}{L_{N}}\int_0^{L_N}\eta(x)dx= 
\frac{\cD_N}{2Na+\cD_N}\approx \frac{b}{a\sqrt{CN}}+o,
\endeq
\ie\, once more  the expected $L^{-1/2}$ behavior. It is possible 
to numerically determine the constant $A\equiv\frac{b}{a\sqrt{C}}$ 
present in  (\ref{4.25}). To this aim, one  first evaluates the $\cD_N$s 
over a large set of $N$values and then one fits the expression 
$A+B/\sqrt{N}$ to the evaluated $\cD_N/\sqrt{N}$ values. Fig. 
\ref{FigB2} shows these values (calculated with 
$b=1,\,c=10^{-2}$ within the range $1\le N\le 10^8$) as well as, 
in red, the resulting fitting curve over the best-fitted $N$ range. 
By  the best-fit one finds that  $A=13.8401$ 
(and $B=0.000095$).\\ 
The FT of $\eta(x)$,  evaluated over the interval centered 
at the origin $O$ and containing $2N$ particle/void pairs 
[$N$  pairs on the left  and $N$ on the right of $O$], reads
\begin{eqnarray}\label{4.26}
&&\teta_{N}(q)=(4/q)\sum_{k=1}^N\Bigl\{ 
\cos[q(L_{k-1}+d_{k}/2)]{\sin(q d_k/2)}- \nonumber\\
&&\quad\quad\quad\quad\cos[q(L_{k-1}+d_{k}+z_k/2)]
{\sin(q z_k/2)}\Bigr\},
\end{eqnarray}
where  we have used the eveness of $\eta(\cdot)$ and put  $L_0=0$. \\
Figure \ref{FigB3}    shows the behavior of $\teta_{N}(q)/L_{N}$ in terms 
of $N$ for some of the typical $q$ values that we choose to consider.
All the panels refer to $N$ values ranging from 1 to $4\times10^7$.  
The shown values of $|\teta_{N}(q)|/\sqrt{L_N}$ refer to $N$s spaced 
by  $10^3$. This could make the real oscillations of the plotted quantity 
wider than it appears from the figures. This should not happen because 
the evaluation was also  performed over the last $10^3$ values of $N$ 
with a unit spacing and the oscillations are of the size present in the 
curve tails.  In each panel, the curve of a given color refers to the $q$ 
value of the same color. All the shown curves indicate that, as $N$ 
becomes sufficiently large,  $|\teta_{N}(q)|/\sqrt{L_N}$ approaches to a 
constant,  the value of which depends on $q$.  In this way,  
relation (\ref{4.2})  appears  numerically to be obeyed. 
The square of  the $|\teta_{N}(q)|/\sqrt{L_N}$ value at the largest 
$N$ value yields a numerical approximation of the scattering intensity. 
Fig. \ref{FigB4} shows the corresponding values in a log-log plot for 
all the $q$ values that we have considered. The resulting shape 
conforms to those usually met in SAS. The oscillations present in 
the tail region are likely related to the fact that most of the particles 
and voids have size close to 2. In fact, the three smallest intensities 
are found at $q=\pi,\,2\pi,\, 4\pi u^{-1}$. These oscillations and 
the small numerical values of the intensities make hard to 
numerically ascertain that sum-rule (\ref{4.2e}) is obeyed.  
Aside from this point numerically not assessed, all the 
other results conform to (\ref{4.2}) so that one can, somewhat  confidently, 
conclude that the $\eta(x)$ defined by 
 (\ref{4.20}) is a SDF. 
 \\ 
\subsubsection*{Conclusions} 
The above analysis  leans upon Wiener results and 
assumptions \bA, \bB\ and \bC\  that are suggested 
by experimental results. It has been shown that assumption \bC\ 
combined with Wiener analysis leads to the mathematical characterization 
of SDFs through statement {\bf S$_2$}. In this way, any $\eta(x)\in \cWp$ 
can be considered a SDF because it generates, {\em via} definitions 
(\ref{2.4}) and (\ref{2.5}), a CF $\gamma(x)$ continuous  throughout 
$R$ and such that  the associated Fourier integral $\sigma(q)$, 
defined by equation (\ref{3.6}), is a non-decreasing and absolutely 
continuous function. Besides, the resulting   $\gamma(x)$ is an  $L_2$ function and the scattering intensity simply is its Fourier transform. It has 
also been shown that the condition ensuring that an  $\eta(x)$ is  
a SDF can be formulated according to statement \bSa, \ie\ 
that $\eta(x)$ must be such that the limit (\ref{4.2}) exists for any $q$ 
and it is such that sum rule (\ref{4.2e}) also is obeyed. \\ 
We are aware that both procedures of defining physical SDFs are, 
on a practical ground, hard to be applied since they are, to a large extent, 
of implicit nature. Therefore they need 
further theoretical exploitation to get  more direct constraints on the $\eta(x)$s eligible to be physical SDFs. 
From this point of view, our analysis only showed that $\eta(x)$ must 
be bounded, have a behavior irregularly oscillating around zero and not to contain contributions of the form $\sum_J A_j\cos(\alpha_j x)$ with the $\{A_j\}$s finite. The further mathematical constraints on $\eta(x)$ that 
ensure   the continuity and the $L_2$ summability of the 
associated $\gamma(x)$ are still unknown even though it is physically 
plausible that they should involve some probability-theory aspects as 
yet poorly defined. In fact,   the M\'ering-Tchoubar models that we 
have numerically analyzed contain some of these elements. The 
corresponding results indicate that the $\cWp$ set is not void as well 
as the necessity of considering  larger $N$ values for the results 
obtained by numerical computations may be fully trusted.

\subsubsection*{Acknowledgmets} 
S.C. is grateful to Professors P. Ciatti, K. Lechner and M. Matone for useful conversations.
\vfill\eject
\subsection*{References}
\begin{description}
\item[\refup{}] Borel, E. (1909). {\em Rend. Circ. Mat. Palermo} {\bf XXVII}, 
247-271.
\item[\refup{}] Chandrasekharan, K. (1980). {\em Classical Fourier Transforms.} Berlin: Springer-Verlag.
\item[\refup{}] Cervellino, A. \&  Ciccariello, S. (2001).  {\em  J. Phys. A: 
Math. Gen.} {\bf 34}, 731-755.
\item[\refup{}] Ciccariello, S. (2017).   {\em  J. Appl. Cryst.}  {\bf 50}, 594-601.
\item[\refup{}]  Ciccariello, S. (2005). {\em Prog. Colloid Polym. Sci.} 
  {\bf  130}, 20-28.
\item[\refup{}] Debye, XYZ., Anderson, H.R.  \&  Brumberger, H. (1957). {\em J. Appl.  Phys.} {\bf 20}, 679-683.
\item[\refup{}] Feigin, L.A. \&  Svergun, D.I. (1987). {\em Structure Analysis
by Small-Angle X-Ray and  Neutron Scattering}, New York: Plenum Press.
\item[\refup{}] Gradshteyn, I.S. \& Ryzhik, I.M. (1980). {\em Table of 
Integrals, Series, and Products}, New York: Academic Press. 
\item[\refup{}] Guinier, A.  (1952). {\em 
X-Ray Diffraction: In Crystals, Imperfect Crystals, and Amorphous Bodies} New York: Dover. 
\item[\refup{}] Guinier, A. \& Fournet, G. (1955). {\em Small-Angle Scattering of X-rays.} New York: John Wiley.
\item[\refup{}] Kolmogorov, A.N. \& Fomin, S.V. (1980) {\em Elementi di teoria delle funzioni e di analisi funzionale}, Mosca: Ediz. MIR. 
\item[\refup{}] Mahler, K. (1927). {\em J. Math. Phys. Mass. Inst. Technology} {\bf 6}, 158-164.
\item[\refup{}] M\'ering, J. \& Tchoubar, D. (1968). {\em J. Appl. Cryst.} {\bf 1}, 153-65.
\item[\refup{}] Morita, T. \& Hiroike, K. (1961). {\em  Prog. Theor. Phys.} 
{\bf 25}, 537-592.
\item[\refup{}] Plancherel, M. (1915). {\em Math. Ann.} {\bf 7}, 315-326.
\item[\refup{}] Porod, G. (1951). {\em Kolloid Z.} {\bf 124}, 83-114.
\item[\refup{}]  Schuster, A. (1906). {\em Proc. Roy. Soc.} {\bf 17}, 136-140.
\item[\refup{}] Wiener, N. (1930). {\em Acta Math.} {\bf 30}, 118-242.
\item[\refup{}] Wiener, N. (1933). {\em The Fourier Integral and Certain of Its 
Applications}, New-York: Dover Pub. Inc.
\end{description}
\end{document}